\documentclass[11pt]{article}
	
	\newcommand{\blind}{0}
	
    \makeatletter
    \renewcommand\section{\@startsection {section}{1}{\z@}%
                                       {-3.5ex \@plus -1ex \@minus -.2ex}%
                                       {2.3ex \@plus.2ex}%
                                       {\normalfont\fontfamily{phv}\fontsize{14}{17}\bfseries}}
    \renewcommand\subsection{\@startsection{subsection}{2}{\z@}%
                                         {-3.25ex\@plus -1ex \@minus -.2ex}%
                                         {1.5ex \@plus .2ex}%
                                         {\normalfont\fontfamily{phv}\fontsize{12}{14}\bfseries}}
    \renewcommand\subsubsection{\@startsection{subsubsection}{3}{\z@}%
                                        {-3.25ex\@plus -1ex \@minus -.2ex}%
                                         {1.5ex \@plus .2ex}%
                                         {\normalfont\normalsize\fontfamily{phv}\fontsize{11}{13}\selectfont}}
    \makeatother
	
	\usepackage{xcolor}
	\usepackage{url} 
	\usepackage{amsthm,amsmath,amsfonts,amssymb,amsbsy}
    \usepackage[authoryear]{natbib}
    \usepackage[colorlinks,citecolor=blue,urlcolor=blue]{hyperref}
    \usepackage{algorithm2e}
    \usepackage{graphicx,psfrag,epsf}
    \usepackage{enumerate}
    \graphicspath{ {./images/} }
    \usepackage{tabularx,booktabs}
    \usepackage{placeins}
    \usepackage{diagbox}
    \usepackage{multirow}
    \usepackage{caption}
    \usepackage[createShortEnv]{proof-at-the-end}
    \usepackage{tikz}
\newcommand*\circled[1]{\tikz[baseline=(char.base)]{
            \node[shape=circle,draw,inner sep=1pt] (char) {#1};}}

    \newcommand\clearrow{\global\let\rowmac\relax}
    \newcolumntype{Y}{>{\centering\arraybackslash}X}
    \theoremstyle{plain}

    \newtheorem{theorem}{Theorem}[section]

    \pgfkeys{/prAtEnd/local custom defaults/.style={
    text link={See proof in Appendix \ref{apd:A}.}
    }
    }
    \theoremstyle{remark}
    \newtheorem{definition}[theorem]{Definition}

    
\begin{document}
		
		\def\spacingset#1{\renewcommand{\baselinestretch}%
			{#1}\small\normalsize} \spacingset{1}
		
		\if0\blind
		{
			\title{ \emph{ \Large Correlated Bayesian Additive Regression Trees with Gaussian Process for Regression Analysis of Dependent Data}}
			\author{Xuetao Lu $^{a*}$ and Robert E. McCulloch $^b$ \\
			$^a$ Department of Biostatistics, The University of Texas MD Anderson Cancer Center \\
            $^b$ School of Mathematical and Statistical Sciences, Arizona State University \\
			$^*$ Author for correspondence: robert.mcculloch@asu.edu}
			\date{}
			\maketitle
		} \fi
		
		\if1\blind
		{

            \title{\bf \emph{IISE Transactions} \LaTeX \ Template}
			\author{Author information is purposely removed for double-blind review}
			
\bigskip
			\bigskip
			\bigskip
			\begin{center}
				{\LARGE\bf \emph{IISE Transactions} \LaTeX \ Template}
			\end{center}
			\medskip
		} \fi
		\bigskip
		
	\begin{abstract}
    Bayesian Additive Regression Trees (BART) has gained widespread popularity, inspiring numerous extensions across diverse applications. However, relatively little attention has been given to modeling dependent data. To fill this gap, we introduce Correlated BART (CBART), which extends BART to account for correlated errors. With a dummy representation, efficient matrix computation was developed for the estimation of CBART. Building on CBART, we propose CBART-GP, a nonparametric regression model that integrates CBART with a Gaussian process (GP) in an additive framework. In CBART-GP, CBART retrieves the true signal of covariates-response relationship, while the GP extracts the dependency structure of residuals. To enable scalable inference of CBART-GP, we develop a two-stage analysis of variance with weighted residuals approach to substantially reduce the computational complexity. Simulation studies demonstrate that CBART-GP not only accurately recovers the true covariate–response relationship but also achieves strong predictive performance. A real-world application further illustrates its practical utility.
	\end{abstract}
			
	\noindent%

	\spacingset{1.5} 

\section{Introduction}\label{sec:intro}

Bayesian Additive Regression Trees (BART) is a powerful nonparametric method for exploring nonlinear relationships and has demonstrated competitive performance relative to widely used machine learning algorithms \citep{Chipman2010bart}. In BART, the regression of a response variable on a vector of covariates is modeled as a sum-of-trees function plus an error term. Specifically, for observations $\{Y,X\}=\{y_i, X_i=(x_{1i},...,x_{pi})\}_{i=1}^n$ of a univariate response and a $p$-dimensional covariate vector, the BART model is defined as:
\begin{equation} \label{eq:bart_iid}
    y_i = \sum_{j=1}^{m}g(X_i;T_j,M_j)+\epsilon_i, \hspace{3mm} \epsilon_i \overset{\text{i.i.d.}}{\sim} N(0,\sigma^2), 
\end{equation}
where $g(X_i;T_j,M_j)$ is a step function, $T_j$ denotes a binary tree consisting of decision rules and terminal nodes, and $M_j = \{\mu_{j1},\ldots,\mu_{jb_j}\}$ represents the mean values associated with the terminal nodes of $T_j$.
A key assumption in model \eqref{eq:bart_iid} is that the errors are independent and identically distributed (i.i.d.) Gaussian noise. To broaden the applicability of BART, several extensions have relaxed this assumption to allow heterogeneous but still independent errors. For instance, \citet{Pratola2020} assumed $\epsilon_i \sim N(0,s(X_i)^2*\sigma^2)$, where $s(x)$ is an estimable function, and introduced a second ensemble of trees to model heteroskedasticity. \citet{George2019} assumed $\epsilon_i \sim N(0,\sigma_i^2)$ and employed a nonparametric Dirichlet process mixture model for the heterogeneous error variances. Comprehensive reviews of other BART extensions can be found in \citet{CompStat} and \citet{Hill2020}. Despite these advances, existing approaches retain the independence assumption, limiting their applicability in the analysis of dependent data. This paper addresses this gap by introducing Correlated BART (CBART), which extends BART to allow correlated errors $\epsilon \sim N(0,\Sigma)$, where $\epsilon=\{\epsilon_i\}_{i=1}^n$ and $\Sigma$ is a general covariance matrix.  

A common class of covariance structures can be defined through a kernel function $\Sigma_{i,j} = K(s_i, s_j \mid \mathbf{\theta})$, where $\mathbf{\theta}$ is a set of parameters, and $s_i$ and $s_j$ are labels (e.g., temporal or spatial points) for random samples $Y(s_i)$ and $Y(s_j)$, which define a Gaussian process (GP) \citep{Rasmussen2005}. The assumption, $Y \sim N(\mu,\Sigma)$, for the GP is widely used in time series study \citep{Roberts2012,BRAHIMBELHOUARI2004} and spatial statistics \citep{krige1951,Cressie1993}. When covariates are introduced, the model takes the form $Y=f(X)+\epsilon$, where for any observed $X=\mathbf{x}$, the conditional expectation $E[Y|\mathbf{x}] = f(\mathbf{x})$ is an estimation of the covariate–response relationship. When the residuals $\mathbf{y}-f(\mathbf{x})$ exhibit dependence, such as spatio-temporal correlation, assuming independent errors $\epsilon$ can cause the estimated covariate–response relationship to be distorted by overfitting the residual dependence (see Section~\ref{sec:compExmp}). To mitigate overfitting and recover the true covariate–response signal, \citet{Stein1991} proposed the regression kriging model, which assumes a linear mean structure $f(X)=X\beta$ and models the error term $\epsilon$ with a spatial dependency defined by the kernel function $K(s_i, s_j \mid \mathbf{\theta})$. However, the linearity assumption on the covariate–response relationship is often inadequate in nonlinear settings.
Since CBART can handle nonlinear relationships while accommodating dependent errors, we replace the linear form $f(X)=X\beta$ with $f(X)=f_{\text{CBART}}(X)$. As in regression kriging, we retain the dependent error assumption and model it using a Gaussian process (GP). Combining CBART with a GP yields our proposed model, CBART-GP, whose structure offers two key advantages for regression analysis with dependent data, such as time series and spatial applications. (1) CBART-GP can recover the true signal of covariate–response relationship with the estimation of $E[Y|\mathbf{x}] = f_{\text{CBART}}(\mathbf{x})$ by filtering out residual dependence. (2) by combining CBART for modeling the covariate–response relationship with a Gaussian process for capturing residual dependency, CBART-GP yields accurate predictions of new responses $\mathbf{y}_{new}$ for new observed covariates $\mathbf{x}_{new}$.

Another major contribution of this paper lies in computational efficiency. First, CBART requires handling a potentially large covariance matrix $\Sigma$, which imposes a substantially heavier computational burden than standard BART. We develop reordering and block-matrix strategies to mitigate this cost. Second, full Bayesian inference for CBART-GP via Markov chain Monte Carlo (MCMC) is infeasible for even moderately sized datasets. To address this, we introduce a two-stage hybrid estimation strategy combining MCMC and maximum likelihood, substantially reducing computational complexity.  

The remainder of this paper is organized as follows. Section~\ref{sec:CBART} introduces the CBART model, including dummy representation, MCMC updating, efficient matrix computations, and an illustrative example demonstrating CBART’s ability to recover the true covariate–response signal. Section~\ref{sec:CbartGP} presents the CBART-GP model, its components, and the proposed two-stage weighted residuals estimation procedure. Section~\ref{sec:experiments} reports simulation studies demonstrating the advantages of CBART-GP. Section~\ref{sec:real_data} applies CBART-GP to a real-world spatial dataset. Section~\ref{sec:disc} concludes with a discussion. The proposed model and algorithms are implemented in the R package \texttt{CbartGP}, available at \url{https://github.com/lxtpvt/CbartGP}.
	

\section{Correlated BART}\label{sec:CBART}

The correlated BART (CBART) with $m$ binary trees can be defined as follows:
\begin{equation} \label{eq:cbart}
Y = \sum_{j=1}^{m}g(X;T_j,M_j)+\epsilon, \hspace{3mm} \epsilon \sim N(0,\Sigma),
\end{equation}
where $Y = \{y_1,\ldots,y_n\}$, $X=\{X_1,\ldots,X_n\}$, \\
$g(X;T_j,M_j)=\{g(X_1;T_j,M_j),\ldots,
g(X_n;T_j,M_j)\}$, $\epsilon = \{\epsilon_1,\ldots,\epsilon_n\}$, and $\Sigma$ is a general covariance matrix.


\subsection{Dummy Representation}\label{dummyRp}

Similar to BART, the building block of CBART is the binary tree. In order to handle the covariance matrix, we propose the concept of dummy representation for a binary tree as follows: 
\begin{definition}[Dummy representation]\label{df:dummy}
\begin{equation} 
	\label{eq:duRp}
    g(X;T,M) = D \mu
\end{equation}
where
\[
    D = \begin{bmatrix}
    d_{11} & d_{12} & ... & d_{1b} \\
    d_{21} & d_{22} & ... & d_{2b} \\
    \vdots & \vdots & \ddots \\
    d_{n1} & d_{n2} & ... & d_{nb}
    \end{bmatrix}, \hspace{5mm}
    \mu = [\mu_1,\mu_2,...,\mu_b]^T.
\]
In the matrix $D$, each row contains zeros in all entries except for the one linked to the bottom node to which the observation is mapped. For example, the $i^{th}$ row in $D$, $[d_{i1},...,d_{i,j-1},d_{i,j},d_{i,j+1},...,d_{in}]=[0,...,0,1,0,...0]$, demonstrates that the $i^{th}$ observation was mapped to the $j^{th}$ bottom node.
\end{definition}

\begin{figure}[t!]
\centering
\includegraphics[width=0.65\textwidth]{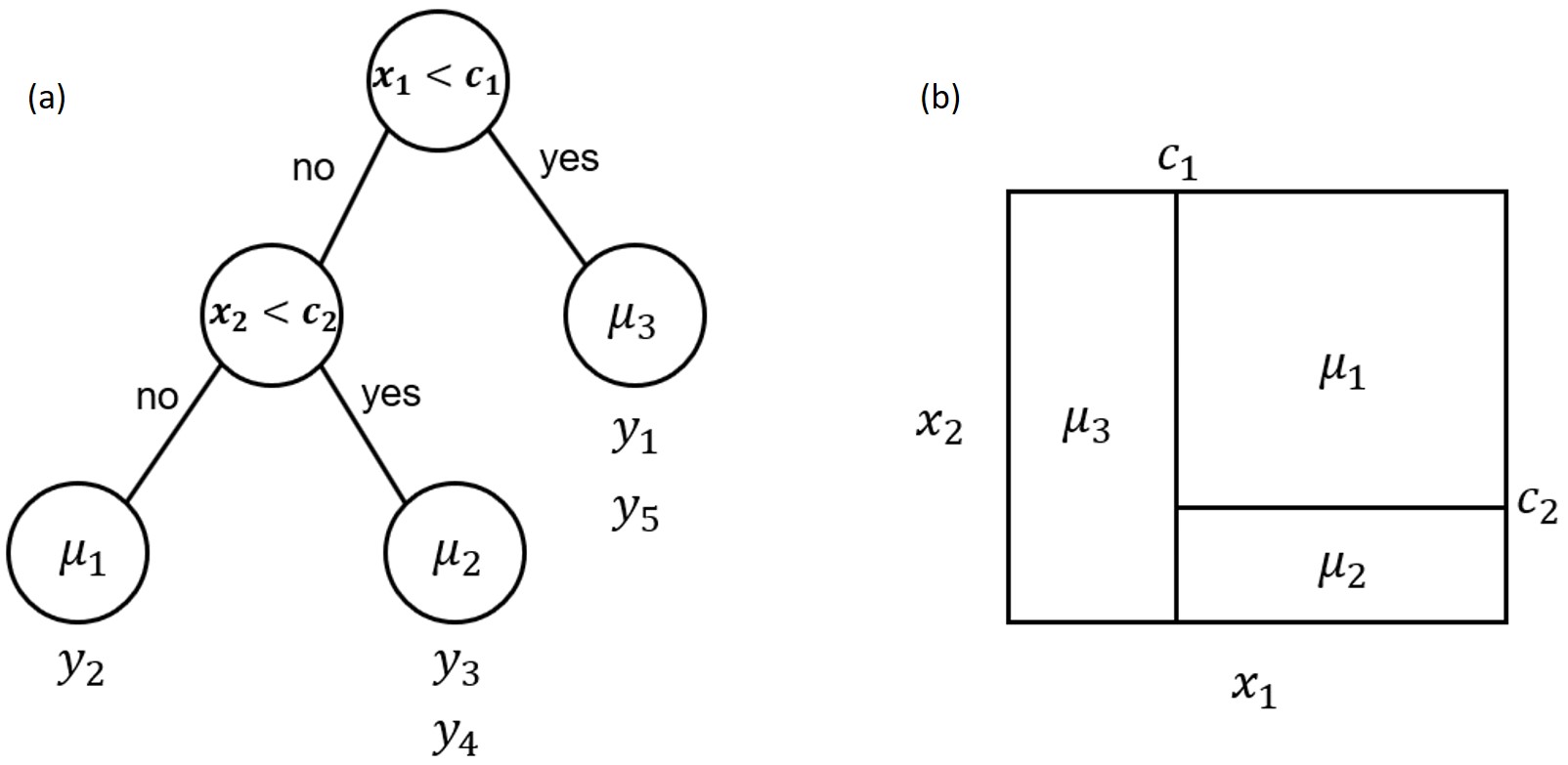}
\caption[Single Tree Model]{An example of the binary tree model, $Y=g(X;T,M)$, where $Y=\{y_1,...,y_5\}$, $X=\{X_1,...,X_5\}$, and $X_i=(x_{1i},x_{2i})$. The tree $T$ including the rules at interior nodes and $M=\{\mu_1,\mu_2,\mu_3\}$ comprising the means of terminal nodes are shown in panel (a). The step function $g(X ; T,M)$ is depicted in panel (b).}
\label{fig:singleTree}
\end{figure}

As an example, the dummy representation of binary tree in Figure \ref{fig:singleTree} is shown as follows. The matrix multiplication serves to map $\{y_2\}$ to $\mu_1$, $\{y_3,y_4\}$ to $\mu_2$, and $\{y_1,y_5\}$ to $\mu_3$.
\begin{equation*}
g(X;T,M) = D\mu =
\begin{bmatrix}
0 & 0 & 1 \\
1 & 0 & 0 \\
0 & 1 & 0 \\
0 & 1 &0 \\
0 & 0 & 1
\end{bmatrix}
\begin{bmatrix}
\mu_1 \\
\mu_2 \\
\mu_3
\end{bmatrix}
\end{equation*}

Based on dummy representation, a binary tree can be denoted as the following matrix form.
\begin{equation} 
	\label{eq:duRpModel}
	Y = g(X;T,M) + \epsilon = D \mu + \epsilon,\hspace{3mm} \epsilon \sim N(0,\Sigma).
\end{equation}

\subsection{Bayesian Backfitting and Markov Chain Monte Carlo}\label{sec:BFMCMC}
To fit the BART model in \eqref{eq:bart_iid}, \cite{Chipman2007,Chipman2010bart} proposed a Bayesian backfitting \citep{Hastie2000} approach, using MCMC to update the pair $(T_j,M_j)$ conditioned on $\sigma^2$ and the remaining trees.  For CBART, we need to deal with $\Sigma$ instead of $\sigma^2$, such that the backfitting reparameterization can be presented as follows:
\begin{equation}\label{eq:singleTree}
R_j = Y - \sum_{k\ne j}g(X;T_k,M_k)=g(X;T_j,M_j)+\epsilon, \hspace{3mm} \epsilon \sim N(0,\Sigma).
\end{equation}
It is easy to check that \eqref{eq:singleTree} denotes an individual binary tree. As the subscript $j$ can take any value in $\{1, \ldots, m\}$, without loss of generality, we simplify the notation \eqref{eq:singleTree} by omitting the subscript hereafter.
\begin{equation}\label{eq:BFsingleTree}
R=g(X;T,M)+\epsilon, \hspace{3mm} \epsilon \sim N(0,\Sigma).
\end{equation}
The key step in BART's MCMC is to update $(T,M)$ given $(R,\Sigma)$. Since we know \\ $p(T,M|R,\Sigma)=p(M|T,R,\Sigma)p(T|R,\Sigma)$, the problem can be converted to the following steps:
\begin{equation} \label{eq:drawT}
T|R,\Sigma
\end{equation}
\begin{equation} \label{eq:drawM}
M|T,R,\Sigma
\end{equation}

\noindent \textbf{(1) Updating $T$ through Metropolis-Hastings algorithm}

\cite{Hugh1998} introduced Metropolis–Hastings algorithm for updating $T$ in \eqref{eq:drawT}. A Markov chain updates the states of the tree in a sequence as follows :
\begin{equation*}
T^0, T^1, T^2,\cdot \cdot \cdot
\end{equation*}
Starting from an initial state $T^0$, the state transition from $T^i$ to $T^{i+1}$, $i=0,1,2,...$, follows two steps:
\begin{itemize}
    \item Given the current state $T^i$, generate a candidate state $T^*$ according to the transition kernel $q(T^i, T^*)$.
    \item Set $T^{i+1} = T^*$ with the probability,
\begin{equation}
\label{eq:M-H-alpha}
\alpha(T^{i+1} = T^*)=min \{ \frac{q(T^*, T^i)}{q(T^i, T^*)}\frac{p(R|X,T^*)p(T^*)}{p(R|X,T^i)p(T^i)},1 \}.
\end{equation}
Otherwise, set $T^{i+1} = T^i$.
\end{itemize}

The terms $q(T^*, T^i), q(T^i, T^*), p(T^*), p(T^i)$ in \eqref{eq:M-H-alpha} are chosen in the same way as for BART. The part that is different between BART and CBART is the marginal likelihood ratio:
\begin{equation}\label{eq:LRimport}
\frac{p(R|X,T^{*})}{p(R|X,T^i)}=\frac{p(R|X,T^{i+1})}{p(R|X,T^i)}.
\end{equation}

Given $\Sigma$, based on the dummy representation of \eqref{eq:duRpModel}, we have that
\begin{equation}
\label{eq:integration}
 p(R|X,T) = p(R|D) = \int p(R|D,\mu) \pi(\mu) \,d\mu,
\end{equation}
where
\begin{equation}
\label{eq:treeRpDist}
 p(R|D,\mu) \sim N(D\mu,\Sigma), \hspace{3mm} \pi(\mu) \sim N(\bar{\mu},Q^{-1}),
\end{equation}
mean $\bar{\mu}$ and precision matrix $Q$ are pre-specified hyperparameters.

\begin{theoremE}[Marginal likelihood ratio]
\label{th:marginal_D}
Given \eqref{eq:integration} and \eqref{eq:treeRpDist}, the marginal likelihood ratio \eqref{eq:LRimport} can be expressed as follows:
\begin{equation}\label{eq:mlr}
\begin{split}
&\frac{p(R|X,T^{i+1})}{p(R|X,T^i)}=\frac{p(R|D^{i+1})}{p(R|D^i)}=\frac{|Q^{i+1}|^{\frac{1}{2}}}{|Q^{i}|^{\frac{1}{2}}} \frac{|Q^i+(D^i)^T\Sigma^{-1}D^i|^{\frac{1}{2}}}{|Q^{i+1}+(D^{i+1})^T\Sigma^{-1}D^{i+1}|^{\frac{1}{2}}} \cdot\\
&exp \{\frac{1}{2}R^T\Sigma^{-1}
[D^{i+1}(Q^{i+1}+(D^{i+1})^T\Sigma^{-1}D^{i+1})^{-1}(D^{i+1})^T-\\
&D^{i}(Q^{i}+(D^{i})^T\Sigma^{-1}D^{i})^{-1}(D^{i})^T]\Sigma^{-1}R\}
\end{split}
\end{equation}
where the superscript $i$ and $i+1$ corresponds to state $i$ and state $i+1$, respectively. 
\end{theoremE}

\begin{proofE}

Given \eqref{eq:integration} and \eqref{eq:treeRpDist}, the product of likelihood and prior can be expressed as follows:
\begin{equation}
\label{eq:lklhPr}
\begin{split}
p(R|D,\mu)\pi(\mu)&= (2\pi)^{-\frac{n}{2}}|\Sigma|^{-\frac{1}{2}}exp\{-\frac{1}{2}(R-D\mu)^T\Sigma^{-1}(R-D\mu)\}*\\
&\hspace{15pt}(2\pi)^{-\frac{b}{2}}|Q|^{\frac{1}{2}}exp\{-\frac{1}{2}(\mu-\bar{\mu})^TQ(\mu-\bar{\mu})\} \\
&= (2\pi)^{-\frac{n+b}{2}}|\Sigma|^{-\frac{1}{2}}|Q|^{\frac{1}{2}}* \\
&\hspace{15pt} exp\{-\frac{1}{2}\underbrace{[(R-D\mu)^T\Sigma^{-1}(R-D\mu)+(\mu-\bar{\mu})^TQ(\mu-\bar{\mu})]}_{(*)}\}.
\end{split}
\end{equation}
where,
\begin{equation*}
\begin{split}
(*)&= R^T\Sigma^{-1}R-2R^T\Sigma^{-1}D\mu+\mu^TD^T\Sigma^{-1}D\mu+\mu^TQ\mu-2\bar{\mu}^TQ\mu+\bar{\mu}^TQ\bar{\mu} \\
&= \mu^T(D^T\Sigma^{-1}D+Q)\mu -2\underline{(R^T\Sigma^{-1}D+\bar{\mu}^TQ)}\mu+ R^T\Sigma^{-1}R+\bar{\mu}^TQ\bar{\mu}.
\end{split}
\end{equation*}

Let's introduce a variable $v$ and consider the quadratic form:
\begin{equation*}
\begin{split}
&\hspace{15pt}(\mu-v)^T(D^T\Sigma^{-1}D+Q)(\mu-v) \\
&=  \mu^T(D^T\Sigma^{-1}D+Q)\mu-2\underline{v^T(D^T\Sigma^{-1}D+Q)}\mu+v^T(D^T\Sigma^{-1}D+Q)v.
\end{split}
\end{equation*}

With the equivalence constraint of above underline coefficients,
$v^T(D^T\Sigma^{-1}D+Q)=R^T\Sigma^{-1}D+\bar{\mu}^TQ$,
we can obtain that
\begin{equation}
 \label{eq:v}
 v = (Q+D^T\Sigma^{-1}D)^{-1}(Q\bar{\mu}+D^T\Sigma^{-1}R).
\end{equation}

Therefore, $(*) = (\mu-v)^T(Q+D^T\Sigma^{-1}D)(\mu-v)+C$,
where $$C=-v^T(Q+D^T\Sigma^{-1}D)v+R^T\Sigma^{-1}R+\bar{\mu}^TQ\bar{\mu}.$$

Then, plug $(*)$ into \eqref{eq:lklhPr}:
\begin{equation*}
\begin{split}
&p(R|D) = \int p(R|D,\mu)p(\mu)d\mu  \\
& = (2\pi)^{-\frac{n+b}{2}}|\Sigma|^{-\frac{1}{2}}|Q|^{\frac{1}{2}} exp\{-\frac{1}{2}C\} \int exp\{ -\frac{1}{2}(\mu-v)^T(Q+D^T\Sigma^{-1}D)(\mu-v)\}d\mu \\
& = (2\pi)^{-\frac{n+b}{2}}|\Sigma|^{-\frac{1}{2}}|Q|^{\frac{1}{2}}exp\{-\frac{1}{2}C\} (2\pi)^{\frac{b}{2}} |Q+D^T\Sigma^{-1}D|^{-\frac{1}{2}} \\
 &\hspace{35pt} \cdot  \int (2\pi)^{-\frac{b}{2}} |Q+D^T\Sigma^{-1}D|^{\frac{1}{2}} exp\{ -\frac{1}{2}(\mu-v)^T(Q+D^T\Sigma^{-1}D)(\mu-v)\}d\mu \\
& = \frac{(2\pi)^{-\frac{n}{2}}|\Sigma|^{-\frac{1}{2}}|Q|^{\frac{1}{2}}}{|Q+D^T\Sigma^{-1}D|^{\frac{1}{2}}} exp\{-\frac{1}{2}C\}
\end{split}
\end{equation*}

After simplification, we can get
\begin{equation*}
p(R|D) = \frac{(2\pi)^{-\frac{n}{2}}|\Sigma|^{-\frac{1}{2}}|Q|^{\frac{1}{2}}}{|Q+D^T\Sigma^{-1}D|^{\frac{1}{2}}} exp\{-\frac{1}{2}(- v^T(Q+D^T\Sigma^{-1}D)v + \bar{\mu}^TQ\bar{\mu} + R^T\Sigma^{-1}R)\},
\end{equation*}
where $v=(Q+D^T\Sigma^{-1}D)^{-1}(Q\bar{\mu}+D^T\Sigma^{-1}R)$.
We can further simplify it by setting $\bar{\mu}=0$.
\begin{equation}
\label{eq:mrglMu0}
p(R|D) = \frac{(2\pi)^{-\frac{n}{2}}|\Sigma|^{-\frac{1}{2}}|Q|^{\frac{1}{2}}}{|Q+D^T\Sigma^{-1}D|^{\frac{1}{2}}} exp\{\frac{1}{2}[R^T\Sigma^{-1}D(Q+D^T\Sigma^{-1}D)^{-1}D^T\Sigma^{-1}R - R^T\Sigma^{-1}R]\}
\end{equation}
By plugging \eqref{eq:mrglMu0} in \eqref{eq:LRimport}, after simplification we can prove that the equation \eqref{eq:mlr} hold.

\end{proofE}

\noindent \textbf{(2) Updating $M$ through the posterior of $\mu$}

Based on the following Proposition \ref{th:posteriorMU}, we can update $M$ in \eqref{eq:drawM} by drawing $\mu$ from its posterior distribution.

\begin{propositionE}[Posterior distribution of $\mu$]
\label{th:posteriorMU}
\begin{equation} 
	\label{eq:postMuPf0}
	p(\mu|R) \sim N(\hspace{3pt}(Q+D^T\Sigma^{-1}D)^{-1}(Q\bar{\mu}+D^T\Sigma^{-1}R)\hspace{3pt},\hspace{3pt}(Q+D^T\Sigma^{-1}D)^{-1}\hspace{3pt})
\end{equation}
Let $\bar{\mu}=0$, it can be simplified to
\begin{equation}
\label{eq:postMu}
p(\mu|R) \sim N((Q+D^T\Sigma^{-1}D)^{-1}D^T\Sigma^{-1}R, (Q+D^T\Sigma^{-1}D)^{-1}).
\end{equation}
\end{propositionE}

\begin{proofE}
Given $D$ and $\Sigma$, the posterior distribution of $\mu$ is proportional to the product of its likelihood and prior as follows:
\begin{equation}
\label{eq:postMuCal}
  p(\mu|R) \propto p(R|\mu)\pi(\mu)=p(R|D,\mu)\pi(\mu),
\end{equation}
where $p(R|D,\mu) \sim N(D\mu,\Sigma)$ and $\pi(\mu) \sim N(\bar{\mu},Q^{-1})$.

Then, conduct the same derivation in \eqref{eq:lklhPr} and \eqref{eq:v}, we can prove that
\begin{equation} 
	\label{eq:postMuPf1}
	p(\mu|R) \sim N(\hspace{3pt}(Q+D^T\Sigma^{-1}D)^{-1}(Q\bar{\mu}+D^T\Sigma^{-1}R)\hspace{3pt},\hspace{3pt}(Q+D^T\Sigma^{-1}D)^{-1}\hspace{3pt}).
\end{equation}
\end{proofE}


\subsection{Efficient Matrix Computations}\label{sec:computational_complexity}
In the MCMC updating of BART, computations in \eqref{eq:mlr} and \eqref{eq:postMu} are efficient because of the assumption of i.i.d. errors, i.e., $\Sigma=\sigma^2I$. However, due to the general form of $\Sigma$, in CBART these computations become intensive as the number of observations $n$ increases. 

For a deeper understanding of the computation complexity, let's set 
\begin{equation}
\label{eq:matA}
A=Q+D^T\Sigma^{-1}D,
\end{equation}
which is a $b\times b$ symmetric matrix and $b$ is the number of bottom nodes.  Given matrix $A$, the computation of $|A|$ and $A^{-1}$ in \eqref{eq:mlr} and \eqref{eq:postMu} require $O(b^3)$ operations. As CBART (same as in BART) prefers small trees, where the number of bottom nodes, b, is typically less than 20, the complexity of $O(b^3)$ is manageable when $A$ is known. However, challenges arise in the calculation of $A$, where $\Sigma^{-1}$ needs $O(n^3)$ operations. For a large number of observations, e.g., $n>10^5$, this computation becomes infeasible for most computers. Fortunately, in some applications such as time series and spatial data analysis, $\Sigma$ exhibits sparsity, i.e., most of its entries are zeros. In such cases, the computation complexity can be reduced through some well designed algorithms/methods which have been extensively studied in both applied mathematics \citep{Wolfgang2015, LIN20114071} and statistics \citep{Datta2016,Andrew2018}. The most favorable outcome is reducing the complexity from $O(n^3)$ to $O(n)$.

The special structure of the matrix $D$ in the dummy representation offers an opportunity to improve the computation efficiency of the marginal likelihood ratio \eqref{eq:mlr}. The idea is to reorder $D$ into a block matrix and apply block computation techniques to enhance efficiency. The definition of reordering is as follows:
\begin{definition}[Reordering]\label{df:reordering}
The matrix $D$ in dummy representation \eqref{eq:duRp} can be reordered to a block matrix $D_P$ using a permutation matrix $P$, 
\begin{equation}
\label{eq:perm}
D = PD_P,\hspace{3mm}P^TD=D_P,
\end{equation}
where
$$
D_P = \begin{bmatrix}
d'_{11} & d'_{12} & ... & d'_{1b} \\
d'_{21} & d'_{22} & ... & d'_{2b} \\
\vdots & \vdots & \ddots \\
d'_{n1} & d'_{n2} & ... & d'_{nb}
\end{bmatrix}, \hspace{2mm}
d'_{ij} = \begin{cases}
0, & i \notin \Omega_j, \\
1, & i \in \Omega_j.
\end{cases}
\hspace{2mm} i=1,...,n; \hspace{2mm}j = 1,...,b,
$$
$\Omega_j$ is the index set of observations that be mapped to $j^{th}$ bottom node.
\end{definition}
Recalling the example shown in Figure \ref{fig:singleTree}, reordering can be demonstrated as follows:
\begin{equation*}
D=
\begin{bmatrix}
0 & 0 & 1 \\
1 & 0 & 0 \\
0 & 1 & 0 \\
0 & 1 &0 \\
0 & 0 & 1
\end{bmatrix}
=PD_P=
\begin{bmatrix}
0 & 0 & 0 & 0 & 1 \\
1 & 0 & 0 & 0 & 0 \\
0 & 1 & 0 & 0 & 0 \\
0 & 0 & 1 & 0 & 0 \\
0 & 0 & 0 & 1 & 0
\end{bmatrix}
\begin{bmatrix}
1 & 0 & 0 \\
0 & 1 & 0 \\
0 & 1 &0 \\
0 & 0 & 1 \\
0 & 0 & 1
\end{bmatrix}.
\end{equation*}

\begin{figure}[t!]
\centering
\includegraphics[width=.75\textwidth]{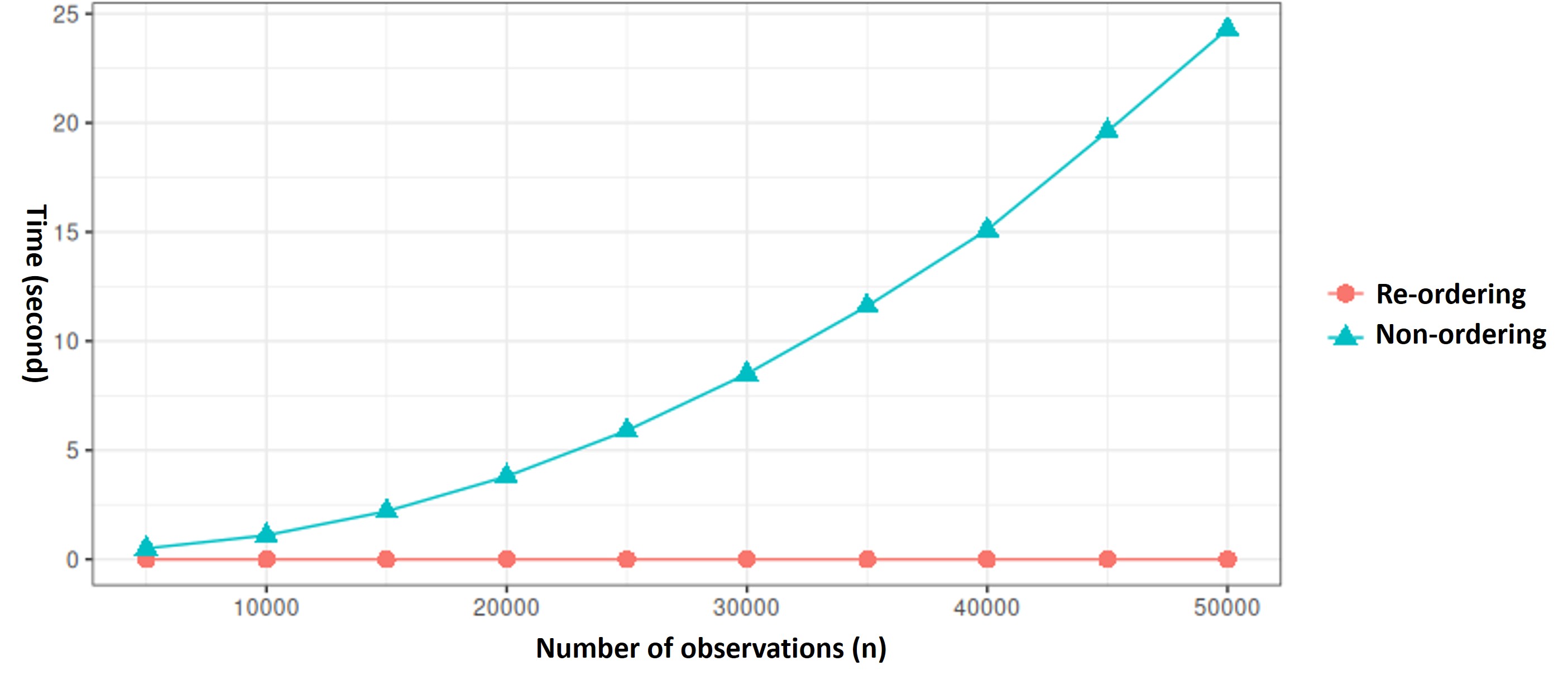}
\caption[Computation efficiency]{The comparison of computational efficiency of re-ordering and non-ordering.}
\label{fig:blkMat}
\end{figure}

For consistency, it is imperative to simultaneously reorder $\{D, R, \Sigma\}$ to $\{D_P, R_P, \Sigma_P\}$ using the same permutation matrix $P$, as follows:
\begin{equation}
\label{eq:perm2}
D = PD_P, \hspace{3mm} R = PR_P, \hspace{3mm} \Sigma = P\Sigma_P P^T.
\end{equation}
Then, we can prove that the marginal likelihood ratio \eqref{eq:mlr} is invariant under reordering by the following Theorem.
\begin{theoremE}[Reordering invariance]
\label{th:reorderInv}
Reordering does not change the value of the marginal likelihood ratio in \eqref{eq:mlr}.
\begin{equation}\label{eq:mlrOrd}
\begin{split}
\frac{p(R|D^{i+1})}{p(R|D^i)}=&\frac{|Q^{i+1}|^{1/2}}{|Q^{i}|^{1/2}} \frac{|Q^i+(D_P^i)^T\Sigma_P^{-1}D_P^i|^{1/2}}{|Q^{i+1}+(D_P^{i+1})^T\Sigma_P^{-1}D_P^{i+1}|^{1/2}} \cdot \\
& exp \{\frac{1}{2} R_P^T\Sigma_P^{-1}[D_P^{i+1}(Q^{i+1}+(D_P^{i+1})^T\Sigma_P^{-1}D_P^{i+1})^{-1}(D_P^{i+1})^T \\
&-D_P^{i}(Q^{i}+(D_P^{i})^T\Sigma_P^{-1}D_P^{i})^{-1}(D_P^{i})^T]\Sigma_P^{-1}R_P\}
\end{split}
\end{equation}
\end{theoremE}

\begin{proofE}
Since $P$ is a permutation matrix, it has the property $P^{-1} = P^T$. With \eqref{eq:perm} and \eqref{eq:perm2}, we can obtain the following:
\begin{equation}
\label{eq:ord1}
\begin{split}
D^T\Sigma^{-1}R &= (PD_P)^T(P\Sigma_PP^T)^{-1}(PR_P) \\
			&= D_P^TP^TP\Sigma_P^{-1}P^TPR_P \\
			&= D_P^T\Sigma_P^{-1}R_P \\
\end{split}
\end{equation}
Given $Q = \tau^{-2}I$, we can get
\begin{equation}
\label{eq:ord2}
Q+D^T\Sigma^{-1}D=Q+(PD_P)^TP\Sigma_P^{-1}P^TPD_P=Q+D_P^T\Sigma_P^{-1}D_P.
\end{equation}
Plug \eqref{eq:perm2}, \eqref{eq:ord1}, \eqref{eq:ord2} into \eqref{eq:mlr}, we can prove the theorem is hold.
\begin{equation*}
\begin{split}
\frac{p(R|D^{i+1})}{p(R|D^i)}&=\frac{|Q^{i+1}|^{\frac{1}{2}}}{|Q^{i}|^{\frac{1}{2}}} \frac{|Q^i+(D^i)^T\Sigma^{-1}D^i|^{\frac{1}{2}}}{|Q^{i+1}+(D^{i+1})^T\Sigma^{-1}D^{i+1}|^{\frac{1}{2}}} \cdot exp \{\frac{1}{2}R^T\Sigma^{-1}[D^{i+1}(Q^{i+1}+\\ 
& (D^{i+1})^T\Sigma^{-1}D^{i+1})^{-1}(D^{i+1})^T-D^{i}(Q^{i}+(D^{i})^T\Sigma^{-1}D^{i})^{-1}(D^{i})^T]\Sigma^{-1}R\}\\
&=\frac{|Q^{i+1}|^{1/2}}{|Q^{i}|^{1/2}} \frac{|Q^i+(D_P^i)^T\Sigma_P^{-1}D_P^i|^{1/2}}{|Q^{i+1}+(D_P^{i+1})^T\Sigma_P^{-1}D_P^{i+1}|^{1/2}} \cdot exp \{\frac{1}{2} R_P^T\Sigma_P^{-1}[D_P^{i+1}(Q^{i+1}+\\
& (D_P^{i+1})^T\Sigma_P^{-1}D_P^{i+1})^{-1}(D_P^{i+1})^T -D_P^{i}(Q^{i}+(D_P^{i})^T\Sigma_P^{-1}D_P^{i})^{-1}(D_P^{i})^T]\Sigma_P^{-1}R_P\}
\end{split}
\end{equation*}
\end{proofE}

The steps for calculating marginal likelihood ratio \eqref{eq:mlrOrd} can be found in Supplementary Materials. Given $\Sigma^{-1}$, by leveraging the block matrix computation technique, the complexity of computing \eqref{eq:mlrOrd} can be reduced from $O(n^2)$ to $max\{O(n),O(b^2)\}$. In addition to the theoretical analysis presented in Supplementary Materials, we conducted a simulation study to compare the computational complexity with and without the reordering technique. In the simulation, we set $b=10$. The results, shown in Table \ref{table:blkMat} and Figure \ref{fig:blkMat}, clearly validate the theoretical findings.

\begin{table}[h]
\caption{Computation time with and without re-ordering. The unit of time is second (s).}
\scriptsize
\begin{tabular}{|c|c|c|c|c|c|c|c|c|c|c|}
\hline
Observations & 5000     & 10000    & 15000    & 20000    & 25000    & 30000    & 35000    & 40000    & 45000    & 50000    \\ \hline
Non-ordering (s)          & 0.05     & 0.16     & 0.51     & 0.83     & 1.20     & 1.67     & 2.27     & 2.93     & 3.66     & 4.48     \\ \hline
Re-ordering (s)          & 2.86e-5 & 3.50e-5 & 2.62e-5 & 2.84e-5 & 3.15e-5 & 3.67e-5 & 4.12e-5 & 4.32e-5 & 4.67e-5 & 4.98e-5 \\ \hline
\end{tabular}
\label{table:blkMat}
\end{table}


\subsection{Example: Recovering the True Covariate–Response Relationship}\label{sec:compExmp}

\begin{figure}[t!]
\centering
\includegraphics[width=1\textwidth]{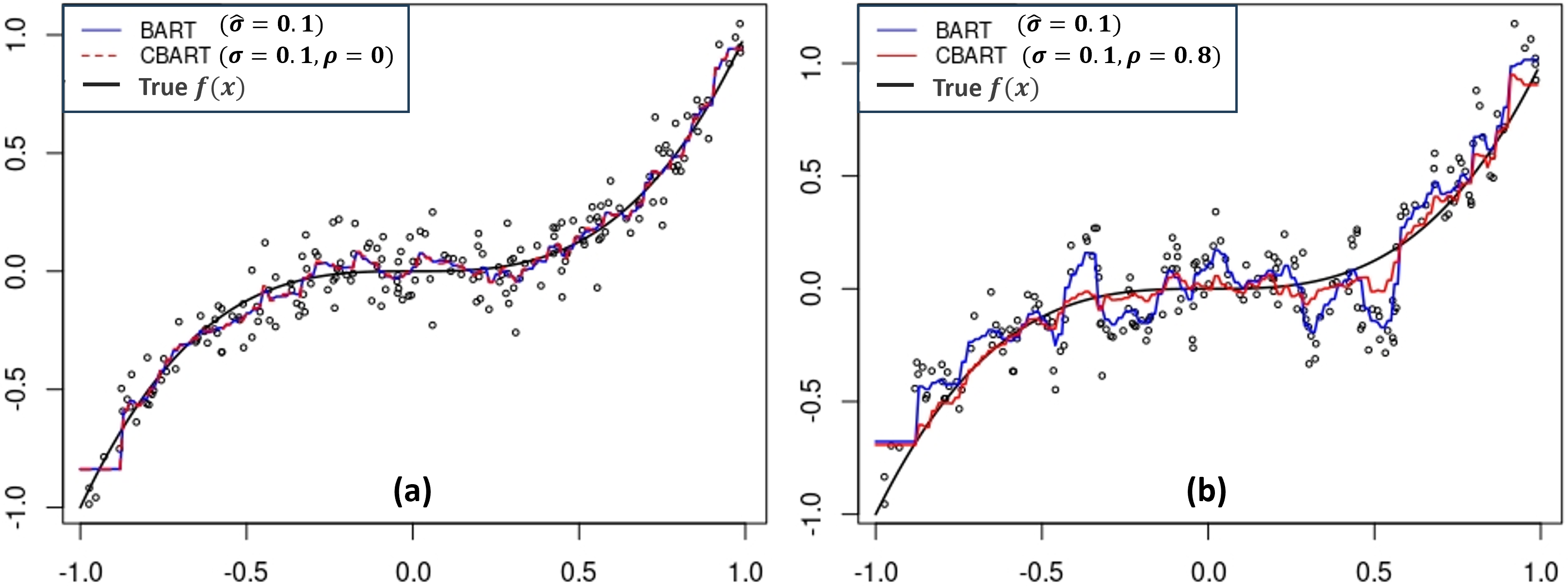}
\caption[Comparing CBART and BART]{Panel (a) illustrates the posterior means of $E[y|x]=f(x)$ estimated from CBART and BART by MCMC when the errors are i.i.d. Panel (b) highlights the distinction between CBART and BART and shows the ability of CBART in recovering the true $f(x)$ when the errors are correlated. Note: $\hat{\sigma}=0.1$ is the BART estimation. $\sigma=0.1$ and $\rho=0$ or $0.8$ are the true parameter values of $\Sigma$ assigned to CBART.}
\label{fig:nbarteg}
\end{figure}

The data for this example is generated from the following model:
\begin{equation}\label{eq:1DExample}
    y_i = f(x_i)+ \eta_i, \hspace{5mm} i\in \{1,...,n\},
\end{equation}
where $n = 200$, $f(x_i) = x_i^3$ and $x_i$ is sampled from a uniform distribution in $(-1,1)$. Let $\boldsymbol{\eta}=\{\eta_1,\ldots,\eta_n\}$ and assume $\boldsymbol{\eta}\sim N(0,\Sigma)$. $\eta_i$ is generated as follows: 
  $$\eta_1=\epsilon_1, \hspace{3mm} \eta_i = \rho*\eta_{i-1}+\epsilon_i,\hspace{3mm} 0\leq\rho<1, \hspace{3mm} i = 2,...,n.$$
  where $\epsilon_i\overset{\text{i.i.d.}}{\sim}N(0,\sigma^2)$, $i = 1,...,n$. We have the matrix form $A\boldsymbol{\eta} = \boldsymbol{\epsilon}$, where
\begin{equation}
\label{eq:example}
A = \begin{bmatrix}
1 & \hdots & \hdots & 0 & 0 \\
 -\rho & 1 & \hdots & 0 & 0 \\
\vdots & \ddots & \ddots & \vdots & \vdots \\
0 & \hdots & \ddots & 1 & 0 \\
0 & \hdots & \hdots & -\rho & 1 
\end{bmatrix}_{n\times n}, \hspace{15pt}
    \boldsymbol{\eta} = \begin{bmatrix}
\eta_1 \\
\vdots \\
\eta_n 
\end{bmatrix}, \hspace{15pt}
    \boldsymbol{\epsilon} = \begin{bmatrix}
\epsilon_1 \\
\vdots \\
\epsilon_n 
\end{bmatrix}.
\end{equation}
Thus, by $\boldsymbol{\eta} = A^{-1}\boldsymbol{\epsilon}$, we can obtain that  $$\Sigma=var(\boldsymbol{\eta})=A^{-1}var(\boldsymbol{\epsilon})[A^{-1}]^T=\sigma^2A^{-1}[A^{-1}]^T, \hspace{5mm} \Sigma^{-1} = \sigma^{-2}A^{T}A.$$

We consider two scenarios: (a) the i.i.d. error case with $\rho=0$, and (b) the dependent error case with $\rho=0.8$. The corresponding datasets are shown in Panels (a) and (b) of Figure~\ref{fig:nbarteg}. For comparison with CBART, we also estimate $f(x)$ using BART under both scenarios. The prior settings are the same as those in \citet{Chipman2010bart}. For CBART, $f(x)$ is estimated using the true covariance matrix $\Sigma$. In practice, however, $\Sigma$ is typically unknown; our method for estimating it is introduced in Section~\ref{sec:two-stage}.

The results are presented in Figure~\ref{fig:nbarteg}. Panel (a) illustrates the i.i.d. error scenario, where CBART and BART yield identical estimates of $f(x)$. Panel (b) shows the correlated error scenario: without accounting for error dependence, BART overfits the residual structure, introducing noise into the estimate of $f(x)$. In contrast, CBART performs much better than BART in recovering the true $f(x)$ by filtering out the influence of dependent errors.


\section{CBART-GP: CBART with Gaussian Process}\label{sec:CbartGP}
Motivated by regression Kriging in spatial statistics, we employ a Gaussian process to capture dependencies in the residuals. Integrating CBART with GP, we propose the model CBART-GP. Unlike regression Kriging, CBART-GP can manage nonlinear regression functions, thus broadening its applicability in practice.

\subsection{Model Specification}\label{sec: ModelSpe}
CBART-GP and its components are outlined as follows:
\begin{equation}\label{eq:CBART-GP}
    y(s_i) = f_{CBART}(\mathbf{x}(s_i)) + \eta(s_i|\mathbf{\theta}), \hspace{3mm} i=1,2,...,n.
\end{equation}
\begin{itemize}
\item $s_i$: the label of observations, e.g., time or location for time series or spatial data;
\item $y(s_i)$: the observed response at $s_i$;
\item $\mathbf{x}(s_i)$: the observed covariates at $s_i$;
\item $\eta(s_i|\mathbf{\theta})$: a Gaussian process with parameters $\mathbf{\theta}$.
\end{itemize}

The Gaussian process $\eta(s_i|\mathbf{\theta})$ can be tailored to different applications by choosing an appropriate parametrization. For example, for time series, it can be parameterized to an AR(p) model:
\begin{equation}\label{eq:CBART-ARp}
   \eta(s_i) = a_1\eta(s_{i-1}) + \ldots + a_p\eta(s_{i-p}) + \epsilon_i, \hspace{3mm} \epsilon_i \sim N(0,\tau^2).
\end{equation}
In this case, $\mathbf{\theta}=\{a_1,\ldots,a_p,\tau^2\}$. We have discussed an example of the AR(1) model in Section \ref{sec:compExmp}.

For spatial data, $\eta(s_i|\mathbf{\theta})$ can be parameterized as follows:
\begin{equation}\label{eq:CBART-Matern}
    \eta(s_i) = z(s_i) + \epsilon_i, \hspace{3mm} \epsilon_i \sim N(0,\tau^2),
\end{equation}
where $z(s_i)$ is a zero mean stationary Gaussian process with Matérn covariance function,
$$Cov(z(s_i),z(s_j))=\sigma^2{\frac {2^{1-\nu }}{\Gamma (\nu )}}{\Bigg (}{\sqrt {2\nu }}{\frac {||z(s_i)-z(s_j)||^2}{\phi }}{\Bigg )}^{\nu }K_{\nu }{\Bigg (}{\sqrt {2\nu }}{\frac {||z(s_i)-z(s_j)||^2}{\phi }}{\Bigg )}.$$
$\Gamma$ represents the gamma function, $K_{\nu}$ denotes a modified Bessel function of the second kind. $\nu$ is a predefined smoothness parameter, with popular choices being $\nu \in \{1/2, 3/2, 5/2\}$. In spatial statistics, $\mathbf{\theta}=\{\sigma^2,\phi,\tau^2\}$ are termed sill, range, and nugget, respectively.

\subsection{Two-stage Analysis of Variance with Weighted Residuals}\label{sec:two-stage}

For the CBART-GP model in \eqref{eq:CBART-GP}, the full Bayesian MCMC sampling strategy can be described as follows:
\begin{equation*}
    \begin{split}
        &(i) \hspace{3mm} \sum^m_{j=1}\{ T_j , M_j\}  \mid \mathbf{\theta} \hspace{1mm} \Leftrightarrow \hspace{1mm} CBART \mid GP \hspace{5mm} \\
        &(ii) \hspace{3mm} \mathbf{\theta} \mid  \sum^m_{j=1}\{ T_j , M_j\} \hspace{1mm} \Leftrightarrow \hspace{1mm}  GP \mid CBART 
    \end{split}
\end{equation*}
where $m$ is the number of trees. 
However, this full Bayesian approach suffers from two inherent limitations: (1) Both CBART and GP are highly flexible due to their nonparametric nature. Under weak prior information on $\mathbf{\theta}$, the MCMC chain mixes poorly and requires a large number of iterations to reach stationary. On the other hand, specifying strong prior information on $\mathbf{\theta}$ is very challenging in practice. (2) As discussed in Section~\ref{sec:computational_complexity}, evaluating $\Sigma^{-1}(\mathbf{\theta})$ in CBART typically requires $O(n^3)$ operations, where $n$ is the sample size. In certain special cases, advanced techniques can reduce this complexity to $O(n)$. Nonetheless, because of limitation (1), the required number of MCMC iterations, say $M$, is typically very large. Thus, the overall computational complexity is at least $O(M \cdot m \cdot n)$ and can reach $O(M \cdot m \cdot n^3)$, where $m$ is the number of trees. This can be prohibitive in practice. For example, with a moderately sized dataset ($n=1000$), assuming $M=10{,}000$ and $m=100$, the complexity is on the order of $O(10^9)$ to $O(10^{15})$, which the computation time is impractically large.

\begin{figure}[t!]
\centering
\includegraphics[width=\textwidth]{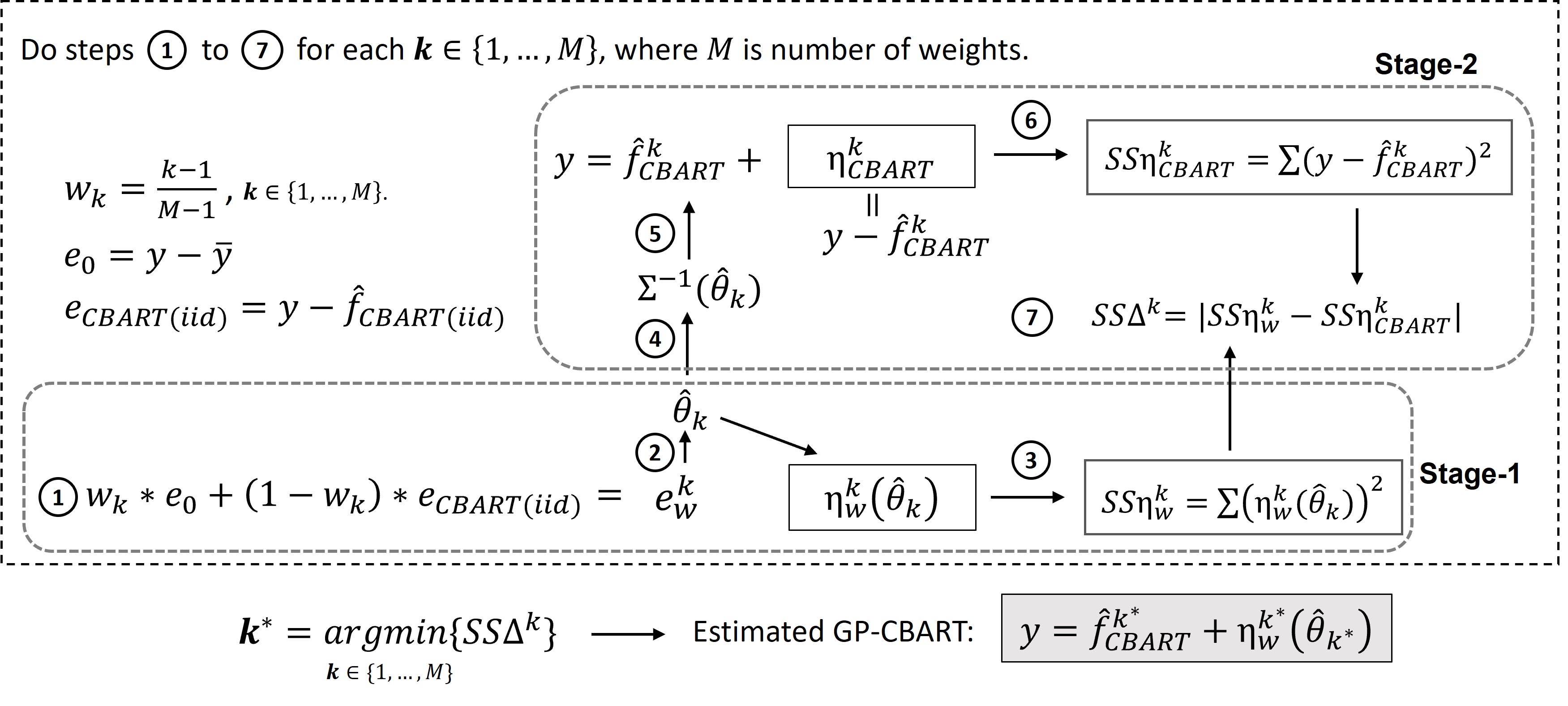}
\caption[Two-stages MLE]{Steps and components of the two-stage analysis of variance with weighted residuals. The number of weights can be set as $M=6$, which performed well in both simulations and real data applications in this paper.}
\label{fig:tsMLE}
\end{figure}

To overcome this computational challenge, we propose a method called two-stage analysis of variance with weighted residuals, which facilitates a hybrid estimation of CBART-GP. 
The steps and components of this approach are illustrated in Figure \ref{fig:tsMLE}. The first stage is designed to estimate the GP, $\eta(\cdot|\mathbf{\theta})$, based on the idea that the dependent structure captured by the GP is contained in the residuals, $e = y - \hat{f}(\mathbf{x})$, where $\hat{f}(\mathbf{x})$ is an estimation of $f(\mathbf{x})$. Two extreme examples are $e_0=y-\bar{y}$ and $e_{CBART(iid)}=y-\hat{f}_{CBART(iid)}$, where $\hat{f}_{CBART(iid)}$ is the estimation from CBART assuming i.i.d. errors, which is equivalent to BART. We regard $e_0$ and $e_{CBART(iid)}$ as representing the maximum and minimum of error dependency to be captured by the GP. Building on this, we introduce weighted residuals, $e^k_w=w_k*e_0+(1-w_k)*e_{CBART(iid)}$, to explore various degrees of dependency within the response. In each scenario, the weight $w_k$ is calculated by $\frac{k-1}{M-1}$, $k=1, \ldots, M$, where $M$ is the number of scenarios. The steps \circled{1} to \circled{3} in the first stage are outlined as follows:
\begin{itemize}
    \item[] Step \circled{1}: model the weighted residuals $e^k_w=w_k*e_0+(1-w_k)*e_{CBART(iid)}$ with the parameterized GP model, $e^k_{w}=\eta_{w}(\theta)$.
    \item[] Step \circled{2}: estimate the MLE $\hat{\theta}_k$ (without $\mathbf{x}$).
    \item[] Step \circled{3}: calculate the variance in $e^k_w$ explained by $\eta^k_{w}(\hat{\theta}_k)$, denoted as $SS\eta^k_{w}=\sum(\eta^k_{w}(\hat{\theta}_k))^2$.
\end{itemize}

\begin{figure}[t!]
\centering
\includegraphics[width=\textwidth]{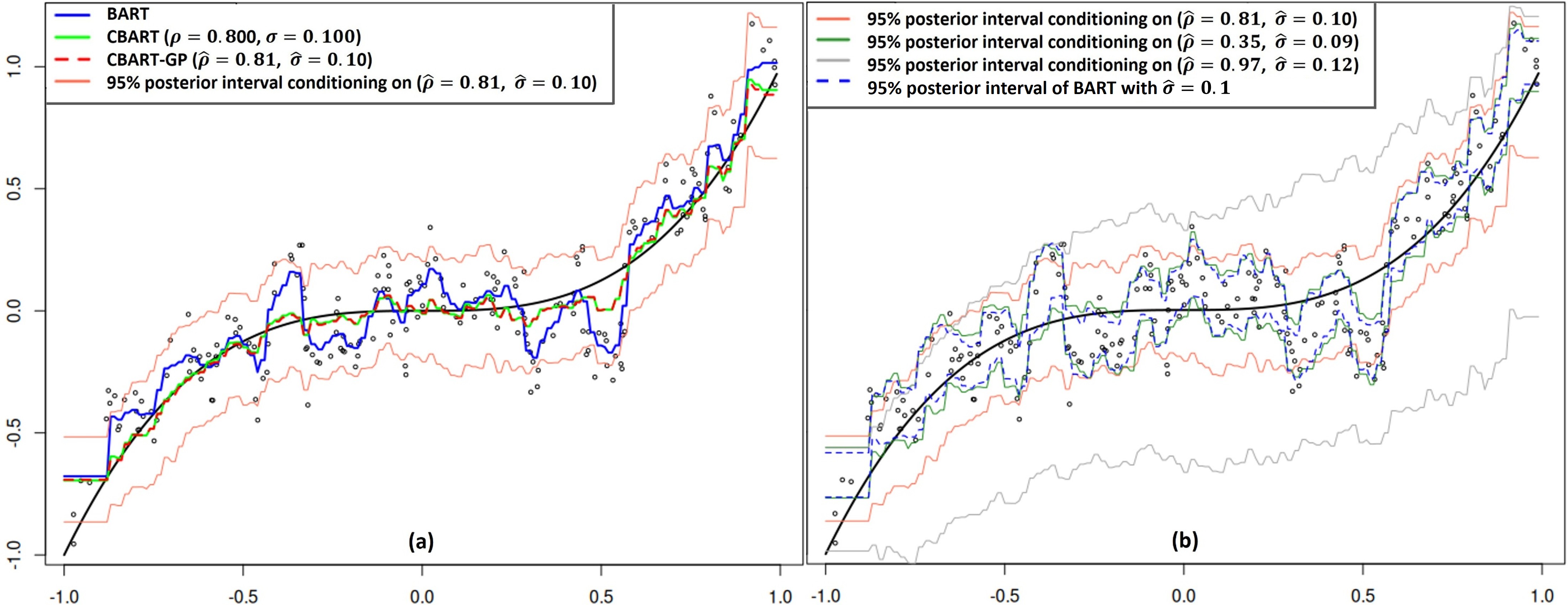}
\caption[Two-stages MLE example]{Illustration of the two-stage analysis of variance with weighted residuals, revisiting the example in Section~\ref{sec:compExmp}. Panel (a): comparison of CBART estimated via the two-stage method with CBART using the true parameters and with BART; the conditional 95\% posterior interval for the CBART estimate is also shown. Panel (b): the 95\% posterior intervals conditioning on different maximum likelihood estimators (MLEs); the 95\% posterior interval for the BART estimate is also shown for comparison.}
\label{fig:tsMLEexample}
\end{figure}

The second stage focuses on estimating CBART and conducting the analysis of variance. Let $\eta^k_{CBART}$ denote the GP model within CBART-GP and $SS\eta^k_{CBART}=\sum(y-\hat{f}^k_{CBART})^2$ represent the variance explained by $\eta^k_{CBART}$. The key idea is to align $\eta^k_{w}(\hat{\theta}_k)$ as closely as possible with $\eta^k_{CBART}$. Their similarity is evaluated through the analysis of variance, aiming to minimize the difference, $SS\Delta^k = |SS\eta^k_{w}-SS\eta^k_{CBART}|$. The steps from \circled{4} to \circled{7} in the second stage are detailed as follows:
\begin{itemize}
    \item[] Step \circled{4}: calculate the inverse covariance matrix $\Sigma^{-1}(\hat{\theta}_k)$.
    \item[] Step \circled{5}: fit the CBART $\hat{f}^k_{CBART}=\hat{f}(\mathbf{x})=\hat{E}[y|\mathbf{x}]$ with $\Sigma^{-1}(\hat{\theta}_k)$.
    \item[] Step \circled{6}: calculate $SS\eta^k_{CBART}=\sum(y-\hat{f}^k_{CBART})^2$ with fitted $\hat{f}^k_{CBART}$.
    \item[] Step \circled{7}: calculate $SS\Delta^k = |SS\eta^k_{e_w}-SS\eta^k_{CBART}|$.
\end{itemize}
After iterating $M$ scenarios of weighted residuals, the optimal CBART-GP model is determined by identifying $k^*$ that corresponds to the minimum $SS\Delta^k$.

The rationale of the two-stage analysis of variance with weighted residuals approach is analogous to that of least square method. In least square, the optimal model (parameters) are those that explain the maximum variation in the data. Similarly, in our two-stage approach, the minimum $SS\Delta^k$ corresponds to the point at which CBART plus GP (i.e., the whole CBART-GP model) explains the maximum variation of the data. 

A larger value of $M$ increases the possibility of reaching the minimum $SS\Delta^k$, but at the cost of higher computational burden. In practice, we recommend discretizing the interval $[0,1]$ into $5\sim 10$ grids for the weights, with $M$ taking values $6\sim 11$. For instance, $M=6$ (5 grids) performs well across all examples and simulations in this paper. Compared to the large number of iterations typically required in a full Bayesian solution (e.g., $M=10{,}000$), our approach substantially reduces computational complexity.

To illustrate the effectiveness of the two-stage analysis of variance with weighted residuals, we revisit the example in Section~\ref{sec:compExmp}. The residual weights are specified as $\{w_1, w_2, w_3, w_4, w_5, w_6\}=\{0,0.2,0.4,0.6,0.8,1\}$, yielding the following values of ${SS\Delta^k}$: $\{1.733, 1.732, \textbf{0.452}, 2.207, 6.221, 10.407\}$. The minimum occurs at $SS\Delta^3 = 0.452$, leading to parameter estimates $\hat{\theta}_3=\{\hat{\rho}_3, \hat{\sigma}_3\}=\{0.81,0.10\}$, which are close to the true values $\theta=\{\rho,\sigma\}=\{0.80,0.10\}$. As shown in Panel (a) of Figure~\ref{fig:tsMLEexample}, the CBART estimate obtained via the two-stage method (red dashed curve) is highly consistent with the estimate using the true parameters (green solid curve).

In addition, we provide a 95\% posterior interval (pink solid curves) for the CBART estimates at each $x$, conditional on the MLEs $\hat{\rho}=0.81$ and $\hat{\sigma}=0.10$. It is important to note that this conditional interval reflects only the uncertainty in CBART estimation given the specific MLEs, and does not account for the variability of the MLEs themselves. Addressing this limitation is a direction for future research. Nevertheless, the two-stage approach enables us to assess the variability of these conditional intervals. Panel (b) of Figure~\ref{fig:tsMLEexample} shows not only the optimal interval corresponding to $w_3=0.4$ (pink solid curves) but also two intervals for extreme cases, $w_1=0$ (dark green solid curves) and $w_6=1$ (gray solid curves), the corresponding MLEs are also reported. In the $w_1=0$ case, the interval closely resembles that of standard BART. Overall, the variation in the 95\% posterior interval lies between the two extremes, with the optimal interval (pink) selected by the two-stage analysis of variance with weighted residuals.


\section{Simulation Studies}\label{sec:experiments}
In this section, we conduct comprehensive simulation studies in both one-dimensional (1-D) and two-dimensional (2-D) settings. We evaluate CBART-GP against alternative models in terms of two key tasks: recovering the true covariate–response relationship and predicting new responses. The comparison models are drawn from three categories: (1) Popular machine learning models: random forest (RF) and support vector machine (SVM). Because the datasets are relatively small ($n=200$ for the 1-D scenario and $n=300$ for the 2-D scenario), large-scale models such as deep neural networks are excluded. (2) BART extensions integrated with Gaussian processes: Treed Gaussian processes (tGP) \citep{GraLee2008}, which combine stationary GPs with the leaf nodes of a single tree to model nonstationary dependence via treed partitioning. GP-BART \citep{MaMurPar2024}, which incorporates GPs into the leaf nodes of BART trees to account for dependence in the data. Both tGP and GP-BART are structurally similar to RF and SVM in that they still assume independence of the errors, which limits their ability to recover the true covariate–response signal. (3) Regression kriging (R-Krig) model.

\subsection{1-D Simulation} \label{sec:sim1}

\begin{figure}[t!]
\centering
\includegraphics[width=.5\textwidth]{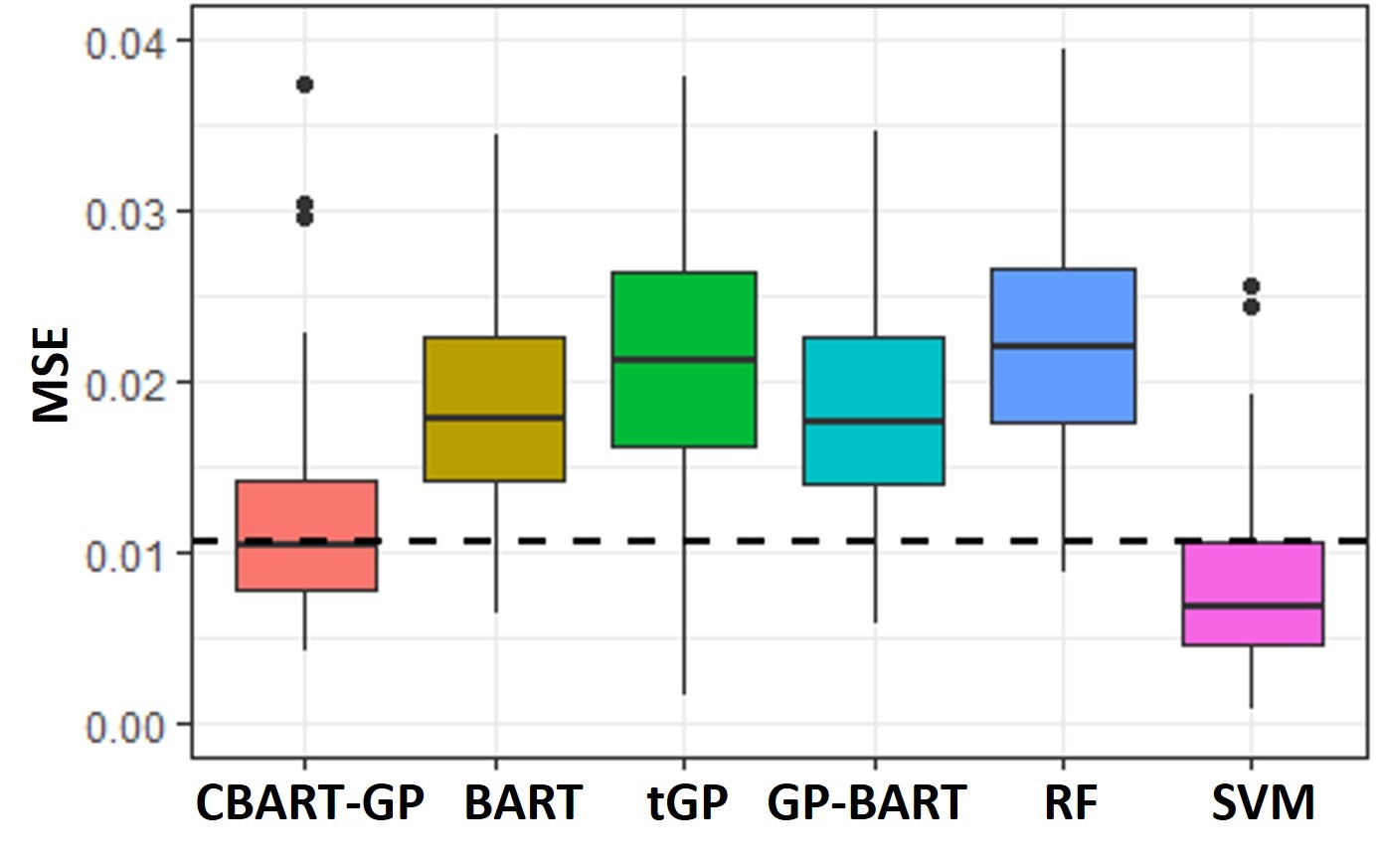}
\caption[Model comparison]{Model comparison in estimating the function $E[y|x]=f(x)$.}
\label{fig:1dcomp}
\end{figure}

In this simulation, we focus on comparing model performance in recovering the covariate–response relationship. We generated 100 datasets using the same settings as in Sections~\ref{sec:compExmp} and \ref{sec:two-stage}. The results, shown in Figure~\ref{fig:1dcomp}, indicate that $\hat{f}_{\text{CBART}}$ outperforms all models except SVM. The superior performance of SVM in this case is due to its inherently stronger smoothing, which aligns well with the true function $f(x)=x^3$, a very smooth relationship. An illustrative example of this effect can be found in Panel (b-4) of Figure~\ref{fig:sim2Ddata} in the next section.

\subsection{2-D Simulation}

Recall the paramterization for spatial data analysis in equation \eqref{eq:CBART-Matern}:
\begin{equation*}\label{eq:sim2}
y(s_i) = f(x(s_i))+ \eta(s_i), \hspace{5mm} \eta(s_i) = z(s_i) + \epsilon_i,
\end{equation*}
where $s_i=(s_{i1},s_{i2})$ represent a spatial point, and $z(s_i)$ denote a Gaussian process $GP(0,C(\cdot,\cdot | \theta))$ characterized by the covariance function $C(\cdot,\cdot | \theta)$. The independent and identically distributed (i.i.d.) errors $\epsilon_i$ follow a normal distribution $\epsilon_i \sim N(0,\tau^2)$.
Unlike the methods described in Section \ref{sec:sim1}, in this spatial data simulation, we aim to achieve two primary goals: (1) estimating the true covariate–response relationship $E[y|x]=f(x)$, and (2) predicting a new response $y^*$ based on $x^*$ and $s^*$. The CBART-GP achieve these goals through: (1) $E[y|x]=\hat{f}_{CBART}(x)$, and (2) $\hat{y}^*=\hat{f}_{CBART}(x^*)+\hat{z}(s^*)$. The performance of CBART-GP is compared with models: BART, GP-BART, RF, SVM, and Regression Kriging (R-Krig). For R-Krig, we have: (1) $E[y|x]=\hat{\beta}x$, and (2) $\hat{y}^*=\hat{\beta}x^*+\hat{z}(s^*)$. For the remaining models, all represented as $\hat{f}$, we have: (1) $E[y|x]=\hat{f}(x)$ for interpretation, and (2) $\hat{y}^*=\hat{f}(x^*, s^*)$ for prediction. The tGP is excluded because it cannot handle cases where $x^*$ and $s^*$ are correlated, a scenario that may occur in practice. The settings for data generation are as follows:

\begin{figure}[t!]
\centering
\includegraphics[width=\textwidth]{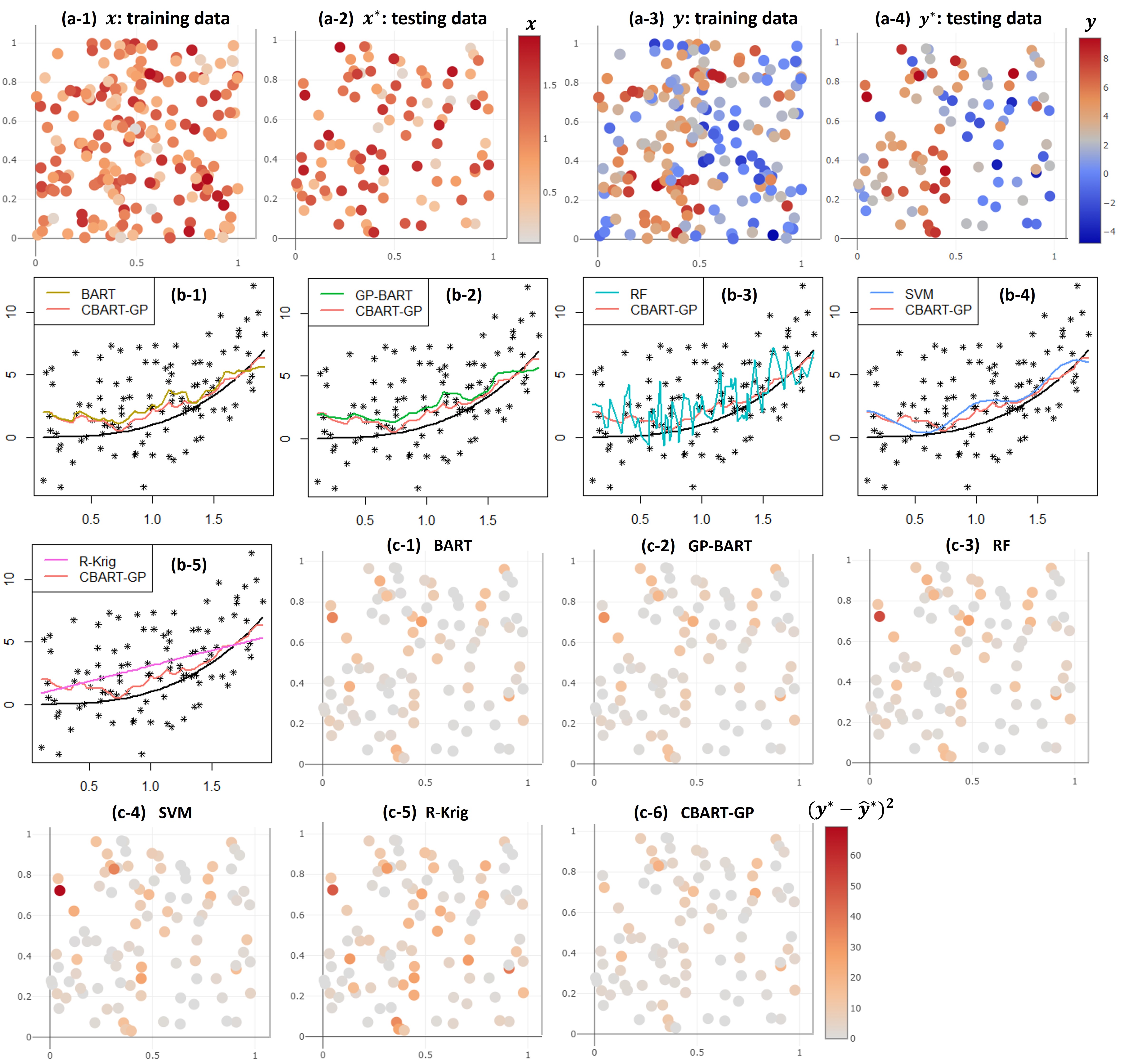}
\caption[2D simulation data]{An example in the simulation of spatial data analysis. Panels (a-1) to (a-4): the training and testing datasets. Panels (b-1) to (b-5): comparison of estimating $E[y|x]=f(x)$. Panels (c-1) to (c-6): comparison of predicting new data $y^*$, which using the criterion of square of error, $(y^*-\hat{y}^*)^2$.}
\label{fig:sim2Ddata}
\end{figure} 

\begin{itemize}
 \item $s_i=(s_{i1},s_{i2})$, where $s_{i1} \perp s_{i2}$ and be generated from the distribution $unif(0,1)$.
 \item $z(s_i) \sim GP(0,C(\cdot,\cdot | \sigma, \phi )), \hspace{2mm} C(s_j,s_k| \sigma, \phi) = \sigma^2exp\{-\frac{d(s_j,s_k)}{\phi} \}$, where $ d(s_j,s_k)$ is the Euclidean distance between the points $s_j$ and $s_k$.
 \item $\epsilon_i \overset{\text{i.i.d}}{\sim} N(0,\tau^2)$.
 \item The true parameters are $\{\sigma^2,\phi,\tau^2\} = \{3,6,1\}$.
 \item The true regression function is $f(x(s_i))= x(s_i)^3$.
 \item $x(s_i)$ is generated to simulate three practical scenarios: (1) $x(s_i)$ is entirely dependent on the spatial point $s_i$: $x(s_i)=s_{i1}+s_{i2}$. (2) $x(s_i)$ is entirely independent of the spatial point $s_i$: $x \sim 2*unif(0,1)$. (3) $x(s_i)$ exhibits a partial dependence and partial independence on the spatial point $s_i$: $x(s_i)=0.5*(s_{i1}+s_{i2})+unif(0,1)$.
\end{itemize}
Similar to the one-dimensional simulation, we generate 100 different datasets under each scenario of $x(s_i)$. Every dataset consists of 200 observations for training and 100 observations for testing. An example dataset of scenario (3) is depicted in panels (a-1) to (a-4) of Figure \ref{fig:sim2Ddata}. 

\renewcommand{\arraystretch}{2.1}
\begin{table}[t]
\caption{ The MSE reduction of CBART-GP comparing to other models in estimating $E[y|x]=f(x)$ and predicting the new data $y^*$.}
\begin{center}
\resizebox{.9\columnwidth}{!}{%
\begin{tabular}{|c|c|c|c|c|c|}
\hline
MSE reduction & BART  & GP-BART & RF & SVM & R-Krig  \\ \hline 
$(\frac{MSE(\hat{f})-MSE(\hat{f}_{CBART})}{MSE(\hat{f})}*100)\%$ & 47.6\% & 50.0\% &  72.6\% & 24.7\% & 65.9\% \\ \hline
$(\frac{MSE(\hat{y}^*)-MSE(\hat{y}^*_{CBART-GP})}{MSE(\hat{y}^*)}*100)\%$ & 21.6\% & 12.8\% &  28.8\% & 35.6\% & 52.6\%   \\ \hline
\end{tabular}
}
\end{center}
\label{table:2DcompEx}
\end{table}

\begin{figure}[t!]
\centering
\includegraphics[width=\textwidth]{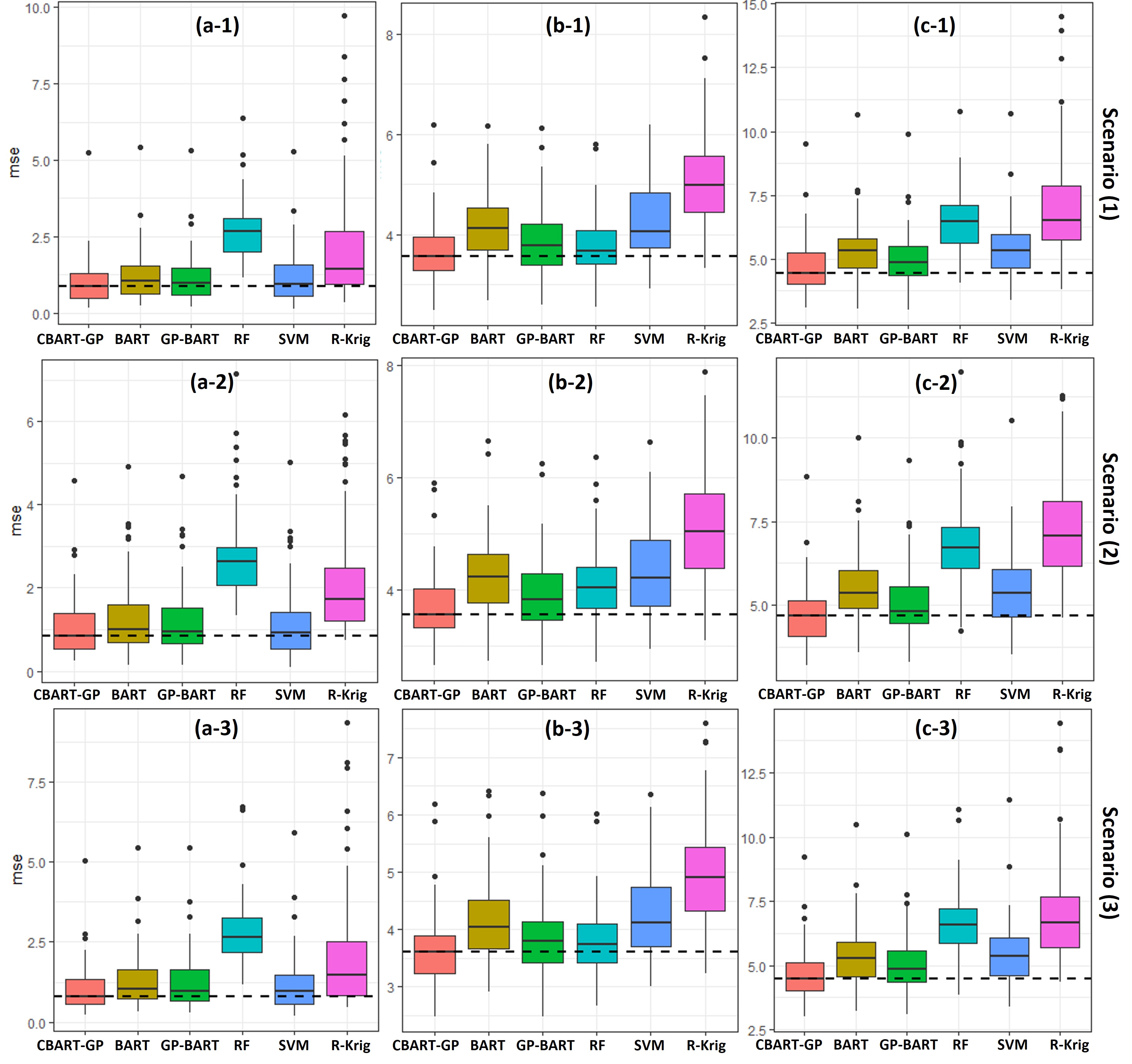}
\caption[Models' comparison]{Simulation results of the spatial data analysis. Panels (a-1) to (a-3): The comparison in estimating $E[y|x]=f(x)$. Panels (b-2) to (b-3): The comparison in predicting new data $y^*$. Panels (c-1) to (c-3): The sum of first two columns, considering the two goals are equally important.}
\label{fig:2dcomp}
\end{figure}

Using the example dataset, we can illustratively examine the models' performance in estimating $E[y|x]=f(x)$ and predicting new data $y^*$. Panels (b-1) to (b-5) in Figure \ref{fig:sim2Ddata} exhibit the results of the former, where the comparisons with CBART-GP are also shown. CBART-GP performs better than other models. For predicting $y^*$, we can compare the models through the square of error, $(y^*-\hat{y}^*)^2$, as shown in panels (c-1) to (c-6). It is not difficult to find that CBART-GP outperforms other models as well. The quantitative measures of CBART-GP's superiority can be found in Table \ref{table:2DcompEx}.

The results of the spatial simulation are depicted in Figure \ref{fig:2dcomp}. Interestingly, the outcomes across three scenarios are consistent, suggesting that spatial dependency of $x$ does not significantly influence the models' performance. Notably, CBART-GP demonstrates best performance in both estimating $E[y|x]=f(x)$ and predicting new data $y^*$. While SVM performs well in the former task due to its smoothness aligning with the true $f(x)$, it performs poorly in the latter task.

\section{Real Data Analysis}
\label{sec:real_data}

\begin{figure}[t!]
\centering
\includegraphics[width=.98\textwidth]{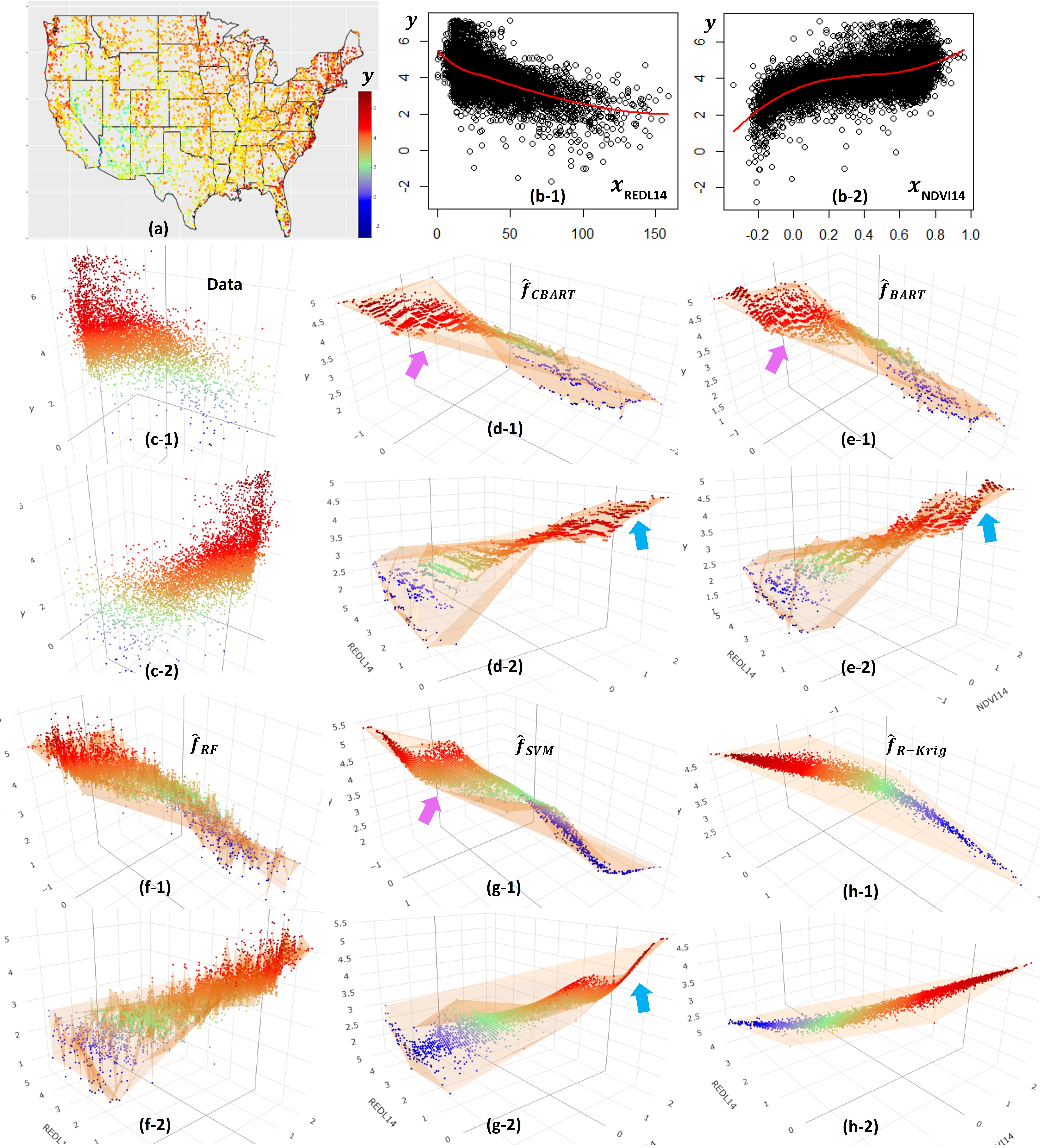}
\caption[The Read Data with Two Environmental Covariates]{Real data analysis. (a) The spatial distribution of the response observations. (b-1,2) Two covariates have nonlinear relationship with response (different angles). (c-1,2) Observations in the 3 dimensional space defined by the response and covariates variables. (d-1,2) CBART's estimation of $f(\mathbf{x})$. (e-1,2) BART's estimation of $f(\mathbf{x})$. (f-1,2) RF's estimation of $f(\mathbf{x})$. (g-1,2) SVM's estimation of $f(\mathbf{x})$. (h-1,2) R-Krig's estimation of $f(\mathbf{x})$. }
\label{fig:realData2v}
\end{figure}

In this section, we apply CBART-GP to a real spatial dataset with 6213 observations, which came from the rapid carbon assessment project initiated by the Natural Resources Conservation Service’s Soil Science Division of the U.S. Department of Agriculture \citep{Wijewardane2016}. The response variable is the soil carbon stock, whose spatial distribution is illustrated in panel (a) of Figure \ref{fig:realData2v}. The dataset contains 31 environmental covariates. Among them, REDL14 (Landsat Band 3 (RED) 2014) and NDVI14 (Normalized Difference Vegetation Index 2014) exhibit nonlinear relationships with the response variable, as depicted in panels (b-1) and (b-2). Panels (c-1) and (c-2) present the observations in the three-dimensional space defined by the response variable and the two covariates. Our goal is to understand the effects of these two covariates on the response, i.e., estimate the true function $E[y|\mathbf{x}]=f(x_{REDL14}, x_{NDVI14})$. We compare the performance of CBART-GP with the models, BART, Random Forest, SVM, and Regression Kriging. tGP and GP-BART are excluded due to their intolerable computation time on this large dataset.

\begin{table}[t!]
\caption{The result of two-stage analysis of variance with weighted residuals.}
\small
\centering
\begin{tabular}{|c|c|c|c|c|c|c|}
\hline
$w_k$ & 0 & 0.2 & 0.4 & 0.6 & 0.8 & \textbf{1}  \\ \hline
$SS\Delta^k$ & 2408 & 2221 & 1874 & 1378 & 717 & \textbf{95}  \\
\hline
\end{tabular}
\label{table:real2D}
\end{table}

Panels (d-1,2) to (h-1,2) display the estimated functions $E[y|\mathbf{x}]=$\\$\hat{f}(x_{REDL14}, x_{NDVI14})$ of different models from two angles. Considering the known nonlinear relationship between covariates and response, the regression Kriging estimation $\hat{f}_{R-Krig}$ shown in panels (h-1,2) likely underestimates the true function. Conversely, the random forest estimation $\hat{f}_{RF}$ in panels (f-1,2) appears to overestimate. The models BART and SVM yield similar estimations ($\hat{f}_{BART}$ and $\hat{f}_{SVM}$) in the areas indicated by the arrows shown in panels (e-1,2), and (f-1,2). The $\hat{f}_{BART}$ and $\hat{f}_{RF}$ appear different from $\hat{f}_{CBART}$ in panels (d-1,2), particularly in the regions indicated by the arrows. The differences were caused by the spatial dependency of the observations. Thus, the following question is whether the spatial dependency is significant or not. This can be evaluated through the estimation of $\theta=\{\sigma^2,\phi,\tau^2\}$. We use the two-stage analysis of variance with weighted residuals method to estimate $\theta$. As shown in Figure \ref{fig:tsMLE}, we need to select the minimal $SS\Delta^k$. In Table \ref{table:real2D}, this corresponds to $w_k=1$, indicating strong spatial dependency in the data. Therefore, we prefer the estimation of $\hat{f}_{CBART}$ over $\hat{f}_{BART}$ or $\hat{f}_{SVM}$. In other word, $\hat{f}_{CBART}$,  as shown in panels (d-1,2), offers a more accurate interpretation of the relationship between the covariates $x_{REDL14}$, $x_{NDVI14}$ and response $y$.

\FloatBarrier

\section{Discussion}
\label{sec:disc}

CBART extends BART by relaxing the i.i.d. error assumption to a general covariance structure, thereby enabling it to accommodate dependent data and recover the true covariate–response relationship. Building on this foundation, CBART-GP offers a novel and powerful regression model for dependent data analysis. Its strength lies in a two-component additive framework that separately captures the effects of observed covariates and the dependence in the response (induced by unobserved covariates). Compared with regression kriging, CBART-GP mitigates bias in estimating the covariate–response relationship by accommodating flexible nonlinear forms when the true function is nonlinear. In contrast to popular nonparametric or machine learning models, CBART-GP avoids overfitting to the covariate–response relationship by effectively isolating residual dependence. Moreover, when predictive accuracy is the main objective, the integration of CBART and GP enables CBART-GP to outperform regression kriging and remain competitive with popular nonparametric and machine learning methods.

Although combining CBART and Gaussian processes offers substantial benefits, it also presents several challenges:
\begin{itemize}
    \item [(1)] Computational burden. As discussed at the beginning of Section~\ref{sec:two-stage}, the computational complexity is at least $O(M \cdot m \cdot n)$ and can reach $O(M \cdot m \cdot n^3)$, where $n$ is the dataset size and $m$ is the number of trees in CBART. In the full Bayesian solution, $M$ denotes the number of MCMC iterations and is typically very large (e.g., $M=10{,}000$). While, in the two-stage analysis of variance with weighted residuals approach, $M$ is a small constant, usually no greater than 11 (e.g., $M=6$). Future research may proceed along two directions: (i) developing computationally efficient algorithms for large $n$, and (ii) the full Bayesian implementations for small datasets, e.g., $n \leq 100$.
    \item [(2)] Posterior intervals. Another challenge lies in constructing posterior intervals for both the CBART estimate $\hat{f}_{CBART}$ and the  parameters covariance matrix in $\Sigma$. For small $n$, the full Bayesian approach may address this issue. However, for moderate or large $n$, new methodologies will be required. A preliminary direction is to investigate the actual coverage of conditional posterior intervals for $\hat{f}_{CBART}$ given the MLEs of GP.
\end{itemize}

\newpage


\appendix
\section{Proofs}\label{apd:A}
\printProofs

\section{Marginal likelihood ratio calculation}\label{apd:B}

Updating $T$ in \eqref{eq:drawT} involves two node operations: birth and death. Birth entails the creation of two child nodes from a leaf node, while death involves removing the two sibling leaf nodes from their parent. For both operations, we need to calculate the marginal likelihood ratio, which includes the following three steps.

\noindent\textbf{(1) Calculate matrix A}

By \eqref{eq:matA} and \eqref{eq:ord2}, we know
\begin{equation*}
A = Q+D_P^T\Sigma_P^{-1}D_P,
\end{equation*}
where $Q = \tau^{-2}I$.

$A$ is a symmetric matrix and can be denoted as follows:
\begin{equation}
\label{eq:matAel}
A= 
\begin{bmatrix}
a_{11}+\tau^{-2} & a_{12} & ... & a_{1b} \\
a_{21} & a_{22}+\tau^{-2} & ... & a_{2b} \\
\vdots & \vdots & \ddots \\
a_{b1} & a_{b2} & ... & a_{bb}+\tau^{-2}
\end{bmatrix},
\end{equation}
where
$$
a_{ji}=a_{ij}=\sum_{h\in \Omega_i}\sum_{l\in \Omega_j}q_{hl}, \hspace{15pt} i\le j, \hspace{15pt} i,j \in \{1,...,b\}.
$$
$\Omega_i$ and $\Omega_j$ are the index sets of observations that are associated with bottom nodes $i$ and $j$; $q_{hl}$ is the entry at the intersection of $h^{th}$ row and $l^{th}$ column in matrix $\Sigma_P^{-1}$ as follows:
$$
\Sigma_P^{-1} = 
\begin{bmatrix}
q_{11} & q_{12} & ... & q_{1n} \\
q_{21} & q_{22} & ... & q_{2n} \\
\vdots & \vdots & \ddots \\
q_{n1} & q_{n2} & ... & q_{nn}
\end{bmatrix}
$$

\noindent\textbf{(2) Calculate the block matrix E}

Pluging $A$ into \eqref{eq:mlrOrd}, we can get
\begin{equation*}
\begin{split}
\frac{p(R|D^{i+1})}{p(R|D^i)}=&\frac{|Q^{i+1}|^{1/2}}{|Q^{i}|^{1/2}} \frac{|A^i|^{1/2}}{|A^{i+1}|^{1/2}}\\
&\cdot exp\{\frac{1}{2}R_P^T\Sigma_P^{-1}\underbrace{[D_P^{i+1}(A^{i+1})^{-1}(D_P^{i+1})^T-D_P^{i}(A^{i})^{-1}(D_P^{i})^T]}_{E}\Sigma_P^{-1}R_P\}.
\end{split}
\end{equation*}

To calculate $E$, we need to consider the birth and death operations separately. Without loss of generality, let's assume that the birth or death operation occurs in the $(i+1)^{th}$ MCMC iteration. 

\textbf{Birth}

In birth operation, the tree has $b$ bottom nodes at $i^{th}$ iteration and $b+1$ bottom nodes at $(i+1)^{th}$ iteration. The corresponding $(A^i)^{-1}$ and $(A^{i+1})^{-1}$ are $b\times b$ and $(b+1)\times (b+1)$ matrices. We can denote them by block matrices as follows:
\begin{equation*}
(A^{i+1})^{-1}=
\begin{bmatrix}
V^{i+1}_{11} & V^{i+1}_{12} \\
V^{i+1}_{21} & V^{i+1}_{22}
\end{bmatrix},
\hspace{15pt}
(A^{i})^{-1}=
\begin{bmatrix}
V^{i}_{11} & v^{i}_{12} \\
v^{i}_{21} & v^{i}_{22}
\end{bmatrix}
\end{equation*}
where, $V^{i+1}_{11}$ and $V^{i}_{11}$ are $(b-1)\times (b-1)$ matrices; $V^{i+1}_{12}=(V^{i+1}_{21})^T$ is $(b-1)\times 2$ matrix; $v^{i}_{12}=v^{i}_{21}$ is a $b-1$ column vector; $v^{i}_{22}$ is a scalar.

We create a matrix
\begin{equation*}
(A^{i})_{ex}^{-1}=
\begin{bmatrix}
V^{i}_{11} & v^{i}_{12} & v^{i}_{12} \\
v^{i}_{21} & v^{i}_{22} & v^{i}_{22} \\
v^{i}_{21} & v^{i}_{22} & v^{i}_{22}.
\end{bmatrix}
\end{equation*}

Let $B=(A^{i+1})^{-1}-(A^{i})_{ex}^{-1}$, then we can get

$$
B=
\begin{bmatrix}
V^{i+1}_{11}-V^{i}_{11} & V^{i+1}_{12}- \begin{bmatrix}v^{i}_{12} & v^{i}_{12}\end{bmatrix} \\
V^{i+1}_{21}-\begin{bmatrix} v^{i}_{21} \\ v^{i}_{21} \end{bmatrix} & V^{i+1}_{22}-\begin{bmatrix} v^{i}_{22} &v^{i}_{22} \\ v^{i}_{22} & v^{i}_{22}\end{bmatrix}
\end{bmatrix}
=\begin{bmatrix}
b_{11} & b_{12} & ... & b_{1(b+1)} \\
b_{21} & b_{22} & ... & b_{2(b+1)} \\
\vdots & \vdots & \ddots \\
b_{(b+1)1} & b_{(b+1)2} & ... & b_{(b+1)(b+1)}.
\end{bmatrix}
$$

\textbf{Death}\newline

Similar to the birth operation, we have $(A^{i})^{-1}$ and $(A^{i+1})^{-1}$ as follows:
\begin{equation*}
(A^{i})^{-1}=
\begin{bmatrix}
V^{i}_{11} & V^{i}_{12} \\
V^{i}_{21} & V^{i}_{22}
\end{bmatrix},
\hspace{15pt}
(A^{i+1})^{-1}=
\begin{bmatrix}
V^{i+1}_{11} & v^{i+1}_{12} \\
v^{i+1}_{21} & v^{i+1}_{22}
\end{bmatrix}
\end{equation*}
where $V^{i+1}_{11}$ and $V^{i}_{11}$ are $(b-2)\times (b-2)$ matrices; $V^{i}_{12}=(V^{i}_{21})^T$ is $(b-2)\times 2$ matrix; $v^{i+1}_{12}=v^{i+1}_{21}$ is a $b-2$ column vector; $v^{i+1}_{22}$ is a scalar. \newline

Create a matrix
\begin{equation*}
(A^{i+1})_{ex}^{-1}=
\begin{bmatrix}
V^{i+1}_{11} & v^{i+1}_{12} & v^{i+1}_{12} \\
v^{i+1}_{21} & v^{i+1}_{22} & v^{i+1}_{22} \\
v^{i+1}_{21} & v^{i+1}_{22} & v^{i+1}_{22}.
\end{bmatrix}
\end{equation*}

In this case, $B=(A^{i+1})_{ex}^{-1}-(A^{i})^{-1}$ as follows:
$$
B=\begin{bmatrix}
V^{i+1}_{11}-V^{i}_{11} & \begin{bmatrix}v^{i+1}_{12} & v^{i+1}_{12}\end{bmatrix} - V^{i}_{12} \\
\begin{bmatrix} v^{i+1}_{21} \\ v^{i+1}_{21} \end{bmatrix}-V^{i}_{21} & \begin{bmatrix} v^{i+1}_{22} &v^{i+1}_{22} \\ v^{i+1}_{22} & v^{i+1}_{22}\end{bmatrix}-V^{i}_{22}
\end{bmatrix}
=\begin{bmatrix}
b_{11} & b_{12} & ... & b_{1b} \\
b_{21} & b_{22} & ... & b_{2b} \\
\vdots & \vdots & \ddots \\
b_{b1} & b_{b2} & ... & b_{bb}
\end{bmatrix}
$$

\textbf{Block matrix E}\newline

$$
E=D_P^{i+1}(A^{i+1})^{-1}(D_P^{i+1})^T-D_P^{i}(A^{i})^{-1}(D_P^{i})^T
=\begin{bmatrix}
E_{11} & E_{12} & ... & E_{1b'} \\
E_{21} & E_{22} & ... & E_{2b'} \\
\vdots & \vdots & \ddots \\
E_{b'1} & E_{b'2} & ... & E_{b'b'}
\end{bmatrix}
$$
where the $b'$ in the subscript of the block matrices $E_{ij}$ is
$$
b' = \begin{cases}
b+1, & \text{Birth}, \\
b, & \text{Death}.
\end{cases}
$$
and, each block $E_{ij}$ has a special form as follows:
$$E_{ij} =E_{ji}^T=b_{ij} \mathbb{J}_{ij}, \hspace{15pt} i\le j, \hspace{15pt} i,j \in \{1,...,b'\}$$
where $b_{ij}$ is the $(i,j)$ entry of matrix $B$ calculated in the birth or death step; $\mathbb{J}_{ij}$ is a card($\Omega_i$) $\times$ card($\Omega_j$) matrix whose entries are all 1. The special form of $\mathbb{J}_{ij}$ reduces the computation load of calculating matrix E from $O(n^2)$ to $O(b^2)$.

\noindent\textbf{(3) Calculate marginal likelihood ratio}

Let's set
$$R_P^T\Sigma_P^{-1}
=\begin{bmatrix}
\omega_{1} & \omega_{2} & \hdots & \omega_{b'}
\end{bmatrix},
\hspace{15pt} \omega_{i} = [\omega_{ij}], \hspace{15pt} j\in n_i
$$
and
$$u = R_P^T\Sigma_P^{-1}E\Sigma_P^{-1}R_P.$$
Then, $u$ can be calculated as follows:
\begin{equation*}
\begin{split}
u &= \begin{bmatrix}
\omega_{1} & \omega_{2} & \hdots & \omega_{b'}
\end{bmatrix}E\begin{bmatrix}
\omega_{1}^T \\ \omega_{2}^T \\ \vdots \\ \omega_{b'}^T
\end{bmatrix}\\
&= \sum_{i=1}^{b'} \sum_{j=1}^{b'} \omega_{i}E_{ij}\omega_{j}^T \\
&= \sum_{i=1}^{b'} \sum_{j=1}^{b'} (\omega_{i}\mathbb{J}_{ij}\omega_{j}^T)b_{ij} \\
&= \sum_{i=1}^{b'} \sum_{j=1}^{b'} [(\sum_{h\in n_i}\omega_{ih})(\sum_{l\in n_j}\omega_{jl})b_{ij}].
\end{split}
\end{equation*}

Finally, we can get the marginal likelihood ratio as follows:

$$
\frac{p(R|D^{i+1})}{p(R|D^i)}=\begin{cases}
\tau^{-1} \frac{|A^i|}{|A^{i+1}|}exp\{\frac{1}{2}u\} \hspace{25pt} Birth \\
\tau \frac{|A^i|}{|A^{i+1}|}exp\{\frac{1}{2}u\} \hspace{35pt} Death.
\end{cases}
$$

The computation load in step (3) is $O(n)$. Thus, the computation complexity of marginal likelihood ratio \eqref{eq:mlrOrd} calculation is $max\{O(n), O(b^2)\}$. A simulation study was conducted to validate this result. In the simulation, we set $b=10$ and $n=\{$ 5000, 10000, 15000, 20000, 25000, 30000, 35000, 40000, 45000, 50000 $\}$. The results are presented in Figure \ref{fig:blkMat} and Table \ref{table:blkMat}, which clearly demonstrate that the reordering technique significantly reduces the computational load.

\newpage


\bibliographystyle{chicago}

\bibliography{GP-CBART}
\end{document}